# Statistical Inference for template-based protein structure prediction

by

**Jian Peng**

Submitted to:

**Toyota Technological Institute at Chicago**

6045 S. Kenwood Ave, Chicago, IL, 60637

For the degree of Doctor of Philosophy in Computer Science

Thesis Committee:

**Jinbo Xu (Thesis Supervisor)**

**David McAllester**

**Daisuke Kihara**

# Statistical Inference for template-based protein structure prediction

by

**Jian Peng**

Submitted to:

**Toyota Technological Institute at Chicago**

6045 S. Kenwood Ave, Chicago, IL, 60637

May 2013

For the degree of Doctor of Philosophy in Computer Science

Thesis Committee:

| | | |
|---|---|---|
| **Jinbo Xu (Thesis Supervisor)** | Signature: | Date: |
| **David McAllester** | Signature: | Date: |
| **Daisuke Kihara** | Signature: | Date: |

# Abstract

Protein structure prediction is one of the most important problems in computational biology. The most successful computational approach, also called template-based modeling, identifies templates with solved crystal structures for the query proteins and constructs three dimensional models based on sequence/structure alignments. Although substantial effort has been made to improve protein sequence alignment, the accuracy of alignments between distantly related proteins is still unsatisfactory. In this thesis, I will introduce a number of statistical machine learning methods to build accurate alignments between a protein sequence and its template structures, especially for proteins having only distantly related templates. For a protein with only one good template, we develop a regression-tree based Conditional Random Fields (CRF) model for pairwise protein sequence/structure alignment. By learning a nonlinear threading scoring function, we are able to leverage the correlation among different sequence and structural features. We also introduce an information-theoretic measure to guide the learning algorithm to better exploit the structural features for low-homology proteins with little evolutionary information in their sequence profile. For a protein with multiple good templates, we design a probabilistic consistency approach to thread the protein to all templates simultaneously. By minimizing the discordance between the pairwise alignments of the protein and templates, we are able to construct a multiple sequence/structure alignment, which leads to better structure predictions than any single-template based prediction. Approaches developed in this dissertation have been implemented as the RaptorX server (http://raptorx.uchicago.edu). The RaptorX server participated the 9th Critical Assessment of Protein Structure Prediction (CASP) experiment. It was ranked the 2nd place among all automated servers and the best for most difficult 50 template-based modeling targets, most of which are low-homology proteins.

# Acknowledgements

My PhD journey has been shared with many people but I don't think the list of names can fit in this single page. First, I would like to thank my advisor Professor Jinbo Xu. His support, motivation, guidance, and knowledge have helped me all the time of my PhD study and research. Without him, I cannot image that I am able to finish all the things I have done in the past six years. Thank you Professor Jinbo, for all the things you have done for me. Also special thanks to my committee Professors David McAllester and Daisuke Kihara for their support and helpful suggestions for my thesis. Second, I would like to thank Professors David McAllester, Nati Srebro, Tamir Hazan, Sham Kakade, Raquel Urtasun, Julia Chuzhoy, Mattias Blume, Greg Shakhnanovich, Joseph Keshet for all the things you have taught me. I also really enjoy the time with you, working on projects, preparing courses, and discussing difficult problems. I would also like to thank my friends at TTI-C, Feng Zhao, Zhiyong Wang, Jianzhu Ma, Sheng Wang, Qingming Tang, Avleen Bijral, Payman Yadollahpour, Taewhan Kim, and Karthik Sridharan. Finally and the most importantly, I would like to thank my wife Jingying for her unending encouragement, support, and love. I thank my parents and parents-in-law for their faith in me and the great help for my life.

# Table of Contents

# List of Figures

# List of Tables

# Chapter 1

# Introduction

Predicting 3D structures of proteins from their amino acid sequences is a grand challenge in computational biology. A number of reasons make protein structure prediction a very difficult problem. The two major reasons are 1) the space of possible protein structure conformations is extremely large, and 2) the physics of protein structural stability is not fully understood. Although the computational prediction methods have achieved significant progress in the last few decades, high-resolution protein structure prediction remains a great challenge.

To evaluate the capability of state-of-the-art prediction methods, a community-wide experiment on the "Critical Assessment of Techniques for Protein Structure Prediction (CASP)" has been organized every other year since 1994 (Moult, Pedersen et al. 1995). The goal of CASP is to get an objective assessment of the current progress of protein structure prediction methods. Participants are asked to predict about 100 soon-to-be-known protein structures in a prediction season. Nowadays, CASPs become the standard benchmark in the community of protein structure prediction. About 80 server teams around the world participated in CASP8 in 2008, CASP9 in 2010 and CASP10 in 2012.

Methodologically, there are two major approaches to predict the structure of a protein, 1) template-based modeling (sometimes also called comparative modeling or protein threading more recently) and 2) *ab initio* modeling/folding. Template-based modeling methods use the previously determined protein crystal structures similar to the query protein to predict the structure for it. This technique is based on the fact that proteins with similar sequences or evolutionary traces tend to have similar structures. Sequence-sequence alignments or sequence-structure alignments are built between the query protein (also called target) and proteins with solved structures (also called templates). Then the structure of the target may be constructed from the alignments, usually by a coordinate-based optimization that utilizes the information from the templates. The accuracy of the alignments will clearly have an effect on the quality of the final 3D predicted structures. Template-based modeling is probably the most successful way in practice so far, despite that it might fail on the targets without good templates. Due to the continuing efforts by structure genomics centers, this method will be likely more effective as the structure space gradually spanned by the Protein Structure Initiative project. The second way, *ab initio* folding, has a more ambitious aim of predicting three-dimensional structures just from the fundamental physics principles. Due to the inaccuracy of physics energy functions, the difficulty on modeling solvent molecules and the lack of efficient sampling techniques, current *ab initio* folding algorithms only achieve limited success on small globular proteins. Recently, approaches such as ROSETTA (Simons, Bonneau et al. 1999; Bonneau, Tsai et al. 2001) and (I-)TASSER (Zhang 2007; Zhang 2008; Zhang 2009), which incorporate template-based information into *ab initio* folding process such that the sampling space is dramatically reduced, are proved to be effective and are ranked among the most successful methods in recent CASP experiments.

My research interests have been focusing on designing novel probabilistic graphical models and statistical methods for template-based protein structure prediction. Under the supervision of Professor Jinbo Xu, I have designed and developed several successful methods (Peng, Bo et al. 2009; Peng and Xu 2009; Xu, Peng et al. 2009; Peng and Xu 2010; Zhao, Peng et al. 2010; Peng and Xu 2011; Peng and Xu 2011; Wang, Zhao et al. 2011; Kallberg, Wang et al. 2012; Ma, Peng et al. 2012), including BoostThreader threading program, which have been implemented in the RaptorX server (available at http://raptorx.uchicago.edu). With BoostThreader and other recently developed methods, our team RaptorX achieved a great success in CASP9 experiment. RaptorX is ranked among the best automatic prediction servers. Notably, it was ranked No.1 in human template-based modeling category, which includes most hard targets; and it was ranked No.2 among all server groups in CASP9. Our method was also voted by CASP community as one of the most innovative methods.

This thesis is organized as follows: in the rest of this Chapter, I will introduce the background of protein structure prediction and template-based modeling/protein threading; in Chapter 2, I will describe BoostThreader, a nonlinear conditional graphical model for protein threading; in Chapter 3, I will present a profile-entropy based measurement to quantify the amount of evolutionary information and to guide protein threading; in Chapter 4, I will present a probabilistic approach for multiple template protein threading. Finally, I will conclude and discuss the future work.

## 1.1 Background of protein sequence and structure

Proteins are molecular machines in the living organisms. Most biological functions in cells, such as catalysis of metabolic reactions, DNA replications, regulation of gene expression, responding to external stimuli and molecular trafficking, are controlled or driven by proteins. Proteins are also the building blocks of tissues. Muscle, hair, nerve, tendon, bone and blood are mostly made up by proteins. Almost half of the non-water mass of a human body is made up of proteins. In molecular biology, the central dogma describes the information flow from DNA to mRNA and finally to proteins. Genes in DNA are transcribed into messenger RNAs and then translated into proteins.

Proteins are macromolecules made of one or multiple chains of amino acids. Most proteins are built from up to 20 canonical amino acids. Except proline, all other 19 amino acids possess a central carbon atom that connects to an amino group, a hydrogen atom, a carboxyl group and a side-chain group. Variable side-chain groups determine the different physicochemical properties of different amino acids, ultimately influencing the different functions of proteins. Amino acids form a linear polymeric chain via the peptide bonds that concatenate the backbone. Each monomer of amino acid in proteins is also called a residue. Thus a protein sequence can be seen as a string of characters from an alphabet with 20 symbols. Although a protein can be simply thought as a string of amino acids, it folds into a unique, compact and stable 3D structure under physiological conditions so to perform its biological functions. The sequence of the protein is also widely believed to contain the full information of its folded structure, which probably corresponds to the minimal free energy of the molecule in solution.

In general, protein structure is described in a hierarchical manner: 1) primary structure which is the sequence of amino acids, 2) secondary structure including irregular loop and two local regular structures – alpha helix and beta sheet, 3) tertiary structure that is formed by packing the local secondary structure elements into one or more compactly connected globular units called domains and 4) quaternary structure which consists of several interacting tertiary structures of the same or different proteins. Most proteins fold into a globular shape in water. There are also proteins that can be folded into filament structures or bundle structures under different physiological conditions.

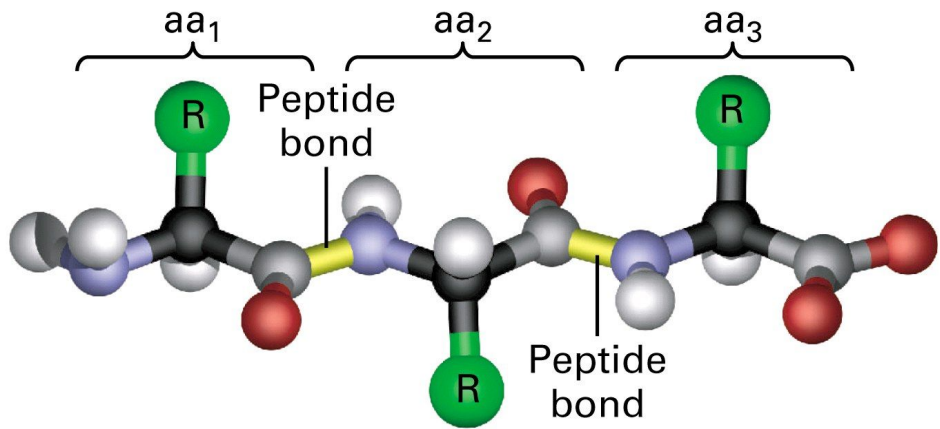

**Figure 1.1.** The backbone of a protein structure.

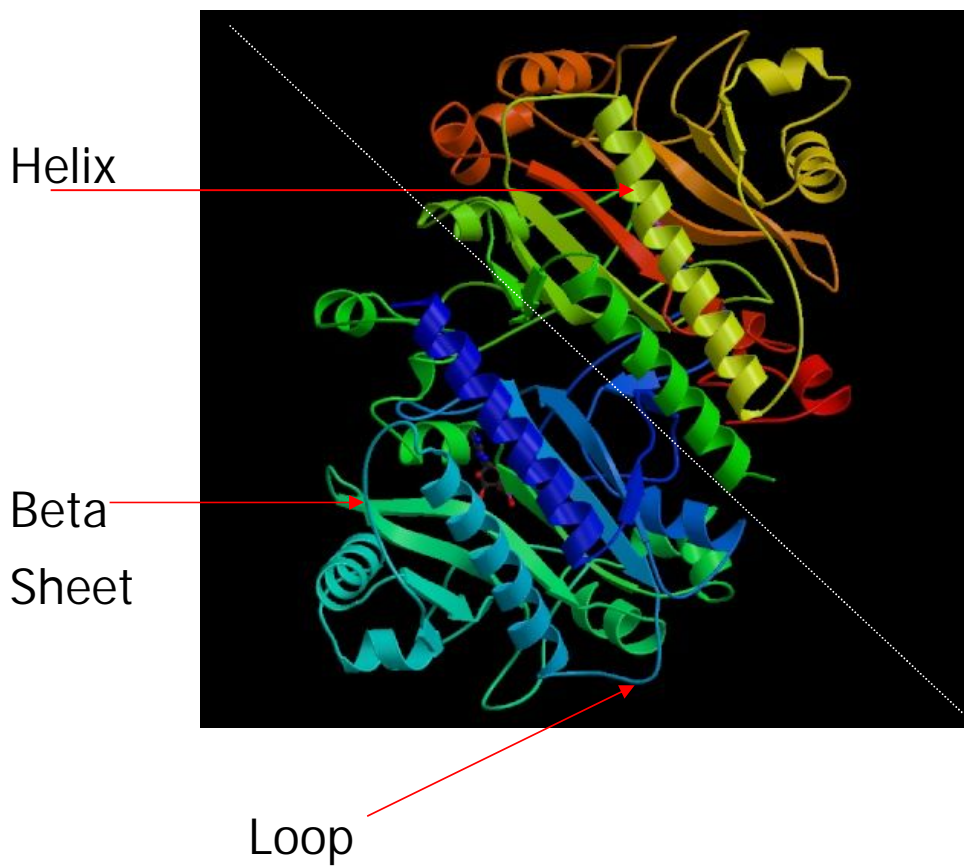

**Figure 1.2.** Alpha helix, beta sheet and loop are three most common secondary structures.

A lot of efforts have been devoted to develop the experimental methods to determine the structure 3D structure of proteins. Two most popular approaches are X-ray crystallography and NMR spectroscopy. Most of the structures deposited at Protein Data Bank (http://www.pdb.org) are solved by X-ray crystallography. To use this method, the protein sample is first prepared, purified and crystallized. Then the crystal is exposed to a beam of X-ray and the diffraction of the X-ray beam is collected and analyzed to determine the distribution of electrons in the protein. Computational methods are then applied to find the coordinate of each atom of the protein according to the electron density map. X-ray crystallography is able to provide very high-resolution detail of the protein structure. Despite of this advantage, X-ray crystallography needs high-quality crystals which can be very difficult to obtain for many proteins. In contrast, NMR spectroscopy doesn't need the crystals. The protein sample can be placed in solution within a strong magnetic field and then be probed with radio waves. Resonances are observed and recorded to provide the proximal information of distant atoms, as well as the local conformations of bonded atoms. The obtained information is then used to build a 3D model of the protein. A key feature of NMR method is that it can provide the protein conformations in solution, similar to the physiological conditions in cell. Nevertheless, NMR method usually has a lower resolution (2 to 4 Angstrom) than X-ray method, and is currently limited to small and medium soluble proteins. Recently, substantial progress has been made on the electron microscopy (EM) techniques. EM is currently applied to determine the structure of large protein complexes. Electron beams are used to image the protein molecules directly. The 3D images are then analyzed to generate 3D electron density maps as the spatial information of the structure. Different from X-ray and NMR methods, EM can only provide the blurb or the approximate shape of the structures with relatively low resolution (8 to 20 Angstrom), but it can deal with large protein complexes. Atomic structures of protein chains or domains, from X-ray crystallography and NMR spectroscopy, are docked together according to the 3D electron density maps from electron microscopy. This strategy has been proved to be effective for multi-molecular structures, such as ribosomes and heat shock protein complexes.

Since it is widely believed that the amino acid sequences encode all the information of the corresponding protein structures (also called Anfinsen's dogma (Anfinsen, Harrington et al. 1955)), it is rationale to ask whether one can predict protein structures directly from their sequences, using the basic laws of physics and chemistry. To achieve this goal, two major issues need to be addressed. First, what is the driving force for protein folding? It is widely believed that a physics-based energy function that guide the protein to be folded into the native conformation with the lowest free energy. However, the exact description of this energy function is still a mystery. To approximate this "correct" energy function from existing observed protein structures, many physics-based or empirical potential functions have been proposed. Physics-based functions, directly derived from the first principles, not only approximate the interactions among all atoms in the protein structure but also try to model the solvent molecules around the protein and their interactions with protein atoms (Roterman, Gibson et al. 1989; Roterman, Lambert et al. 1989; Kini and Evans 1992; Hsieh and Luo 2004). Due to the substantial degrees of freedom, the efficiency of computation and simulation of physics-based energy functions is far from satisfactory. Statistical potential functions, on the other hand, try to approximate the intra-molecular and inter-molecular interactions in the coarse-grained or implicit ways and the parameters associated are directly calculated from a database of solved structures. Statistical potential functions are usually easier to calculate but at a cost of the low-resolution accuracy.

The second issue is how to develop algorithms to enumerate the protein structures. Due to the astronomical number of conformations for a protein, search algorithms which explore the conformation space in efficient ways are another essential component to solve this *ab initio* protein folding problem (Hart and Istrail 1997; Berger and Leighton 1998). These algorithms are also associated with the energy functions used for search. An accurate and smooth energy function will make the search much easier to find the lowest free energy state. However, this is not the case for almost all energy functions that have been proposed. The energy landscape of a protein can be highly "rugged" (Leopold, Montal et al. 1992). The landscape consists of numerous local minima such that a protein can be easily trapped inside with a partially folded conformation. This makes the design of algorithms, which can jump out these local minima, quite difficult.

Despite the limited success of *ab initio* folding in protein structure prediction, template-based modeling, which utilizes the concept of pattern recognition to predict the structures of new proteins from previously determined structures, has been shown to be successful in practice. In the rest of this chapter, we will focus on the basics and methods in template-based modeling.

## 1.2 Template-based modeling for structure prediction

Template-based modeling uses previously solved 3D structures as starting points, or templates to predict the structures of new protein sequences. The rationale of this method is based on the important observation that there is a limited set of tertiary structural folds existing in nature. There are only around 1,500 to 2,000 distinct protein folds found in PDB, while there are many millions of different proteins (Lo Conte, Ailey et al. 2000). Current PDB may contain all templates for single-domain proteins according to the seminal studies in Zhang and Skolnick (Zhang and Skolnick 2005). This implies that the structures of many new proteins can be predicted using template-based methods.

Figure 1.3 demonstrates the basic paradigm of template-based modeling. For any query protein, pairwise alignments to all templates are performed to generate sequence-template alignment and the best template is picked to build the 3D structure of the query protein. After the alignments are built with the best templates, the spatial information from these templates is then used to predict the backbone of the query protein. This step is usually done by MODELLER (Fiser and Sali 2003) or ROSETTA (Rohl, Strauss et al. 2004), which can sample and optimize the 3D conformation to satisfy the geometric constraints extracted from the templates. Finally, the side-chains of residues are added onto the backbone.

The quality of the prediction is mainly determined by the accuracy of the alignment between query target and templates. There are three major components of a protein alignment program, including 1) alignment scoring function which is designed to measure the structural/evolutionary fitness between the residues of the query protein and the residues on templates; 2) a protein alignment algorithm that maximize the scoring function to generate optimal alignment between the query sequence and the templates; and 3) a template selection algorithm which is used to choose the most probable templates. Protein threading is an alignment program which is designed to model the proteins with same fold as known templates even without closely homologous similarity on their sequences. In contrast, homology modeling can only detect the templates with closely-related sequences, which are accurate enough for easy targets. The rest of this thesis will focus on improving methodology for protein threading.

Inherently, protein threading or alignment is a "pattern matching" problem with the goal to find similar patches between two proteins. To measure the similarity between two protein sequences, various measures have been proposed. For closely related homologous sequences (i.e. sequences with high sequence identity), BLOcks of Amino Acid SUbstitution Matrix (BLOSUM) (Henikoff and Henikoff 1992) and other substitution scores have been widely used. Each score represents the fitness between two amino acids. These scores are derived from structural alignments of proteins within their families to capture the statistical preference by substituting one amino acid by another. Since they are constructed from highly similar proteins, however, they are not sensitive enough to build accurate alignments for distantly related protein pairs, which is a major challenge for template-based protein structure prediction.

To capture the similarity between distantly-related proteins, one should exploit extra information instead of only using the amino acid sequences for alignments. Sequence profile of a protein, built on the multiple sequence alignment with sequence homologs, carries extra evolutionary information than its amino acid sequence alone. Represented as a matrix that records the preference of amino acids at all sequence positions, sequence profile is calculated from the position-specific frequencies of the amino acids showing in the evolution. As a result, sequence profiles can capture the local residue conservation and linear motifs shared within a protein family or a superfamily during evolution. Since the protein structures are more conserved than the sequences, each column of sequence profile indeed encodes the functional or structural constraints of that residue of the protein. Taking this information into account, the similarity scores, which can be a distance measure between two matrices, between sequence profiles of two proteins are much more sensitive than BLOSUM scores on detecting distantly-related homologous relationship. Sequence profiles have also been shown extremely effective on predicting many protein structural features, such as secondary structure, solvent accessibility and torsion angles. Thus, almost all existing protein threading methods takes sequence profiles as their input. The accuracy of protein threading thus is heavily relied on the quality of sequence profiles.

Homology search programs are usually used to find evolutionarily related sequences (e.g. sequences belong to the same family or super family), generate multiple sequence alignments and construct the sequence profile for a protein. The most widely used program is probably PSI-BLAST (Altschul, Madden et al. 1997). Different from pure-sequence-based BLASTp (Altschul, Gish et al. 1990), PSI-BLAST iteratively builds a sequence profile called the position specific substitution matrix (PSSM) from the multiple sequence alignment of detected homologous sequences. The PSSM is then used to further search the sequence database for other homologous sequences and is updated with these newly found sequences for the subsequent searching iterations. Besides PSI-BLAST, many other programs, including HMMER (Krogh, Brown et al. 1994; Finn, Clements et al. 2011) and HHBLITS (Remmert, Biegert et al. 2012), also use profile hidden Markov models to construct the sequence profiles from the multiple sequence alignment of detected sequences by sequence search programs.

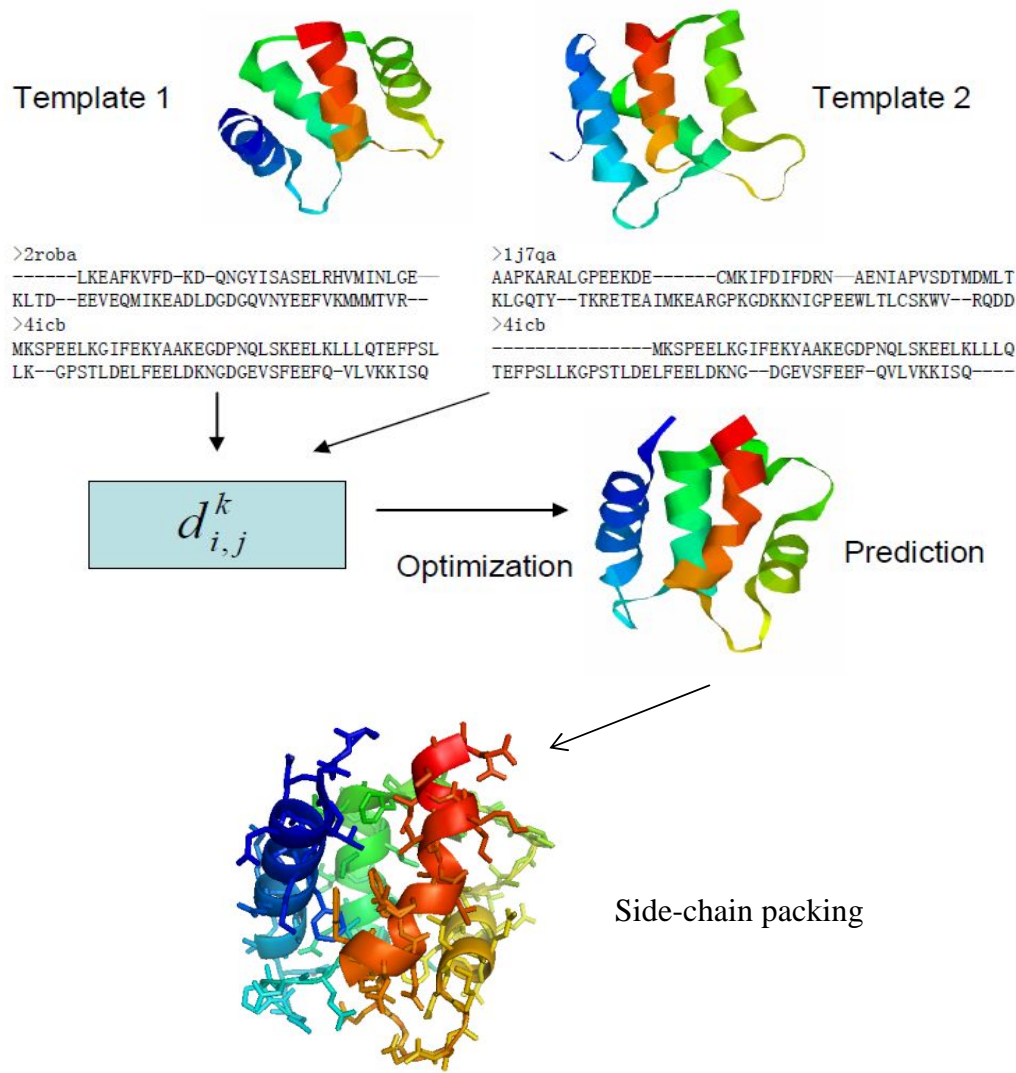

**Figure 1.3.** Template-based modeling. 1) Best templates are identified. 2) The alignment between query target and the templates are generated. 3) A backbone structure is optimized according to the spatial information from templates. 4) Side-chains are added on to the predicted backbone.

Although evolutionary information carved in the sequence profiles provides very precisely homologous similarity between two closely or moderately related proteins, structural similarity seems contribute independent but important signals to build reliable alignment for two remotely related proteins. Since the accuracy of protein threading is heavily relied on the quality of sequence profiles, structural features are especially useful when the profiles cannot provide enough information to build reliable alignments.

In many cases, such as many difficult targets appeared in recent several CASPs, PSI-BLAST failed to find a sufficient number of homologous sequences. As a result, the sequence profiles are highly sparse, limiting the capability of threading program to find the correct templates. To alleviate this issue, structural information has been intensively exploited in this situation. Besides secondary structure prediction, researchers have also developed numerous programs to predict other structural features, such as solvent accessibility and disordered state, directly from amino acid sequences or sequence profiles. Although not perfect, some of them can be predicted quite accurately. For instance, most programs for secondary structure prediction, which assigns helix/beta/loop labels to each residue, can predict with ~80% accuracy (McGuffin, Bryson et al. 2000). If the predicted secondary structure elements are aligned approximately correct with the true secondary structures of the template, the potential search space of possible alignments would be reduced substantially. By taking into account the compatibility of these predicted structural features of query sequences with the crystal structural features of templates, threading programs are able to generate more reliable alignments when the signals within sequence profiles are sparse or noisy. In most threading programs, the compatibility of structural features is usually implemented as one term of the scoring function, thus imposing extra structural constraints that could help guide the alignment with the evolutionary information. Figure 1.5 shows a flowchart for protein threading.

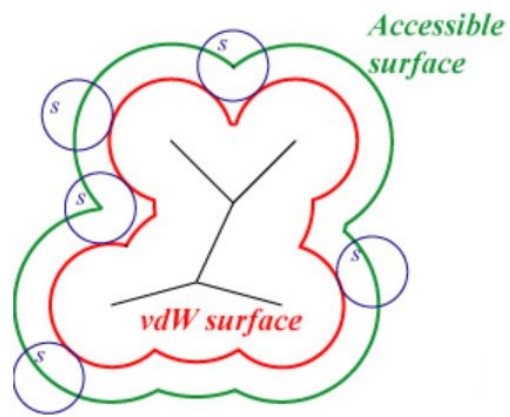

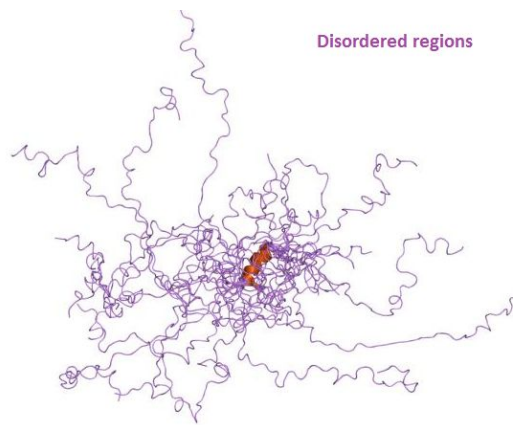

**Figure 1.4.** Examples of structural features used in protein threading.

There have been decade-long efforts made by researchers to develop effective programs for protein threading. Representative methods include HHpred (Soding, Biegert et al. 2005; Hildebrand, Remmert et al. 2009), RAPTOR (Xu, Li et al. 2003), MUSTER (Wu and Zhang 2008) and Sparks (Zhou and Zhou 2005). HHpred belongs to the category of profile-profile alignment algorithms. Profile-based algorithms compare the query protein and the template protein through their sequence profiles. HHpred extends the sequence profile concept with hidden Markov models which include the position-specific indels frequencies. Besides sequence profiles, RAPTOR, MUSTER and Sparks also exploit the structural features to build alignments, such as those mentioned above. Structural features are shown to be effective especially for proteins which have few homologs detectable by sequence or profile-only search methods. New versions of HHpred also incorporate some structural features, such as secondary structures, to improve its power on detecting distantly related templates. The scoring functions of these approaches usually additively combine the similarity of homology information from sequence profile, and the structural fitness of aligning query sequence to local environment on template.

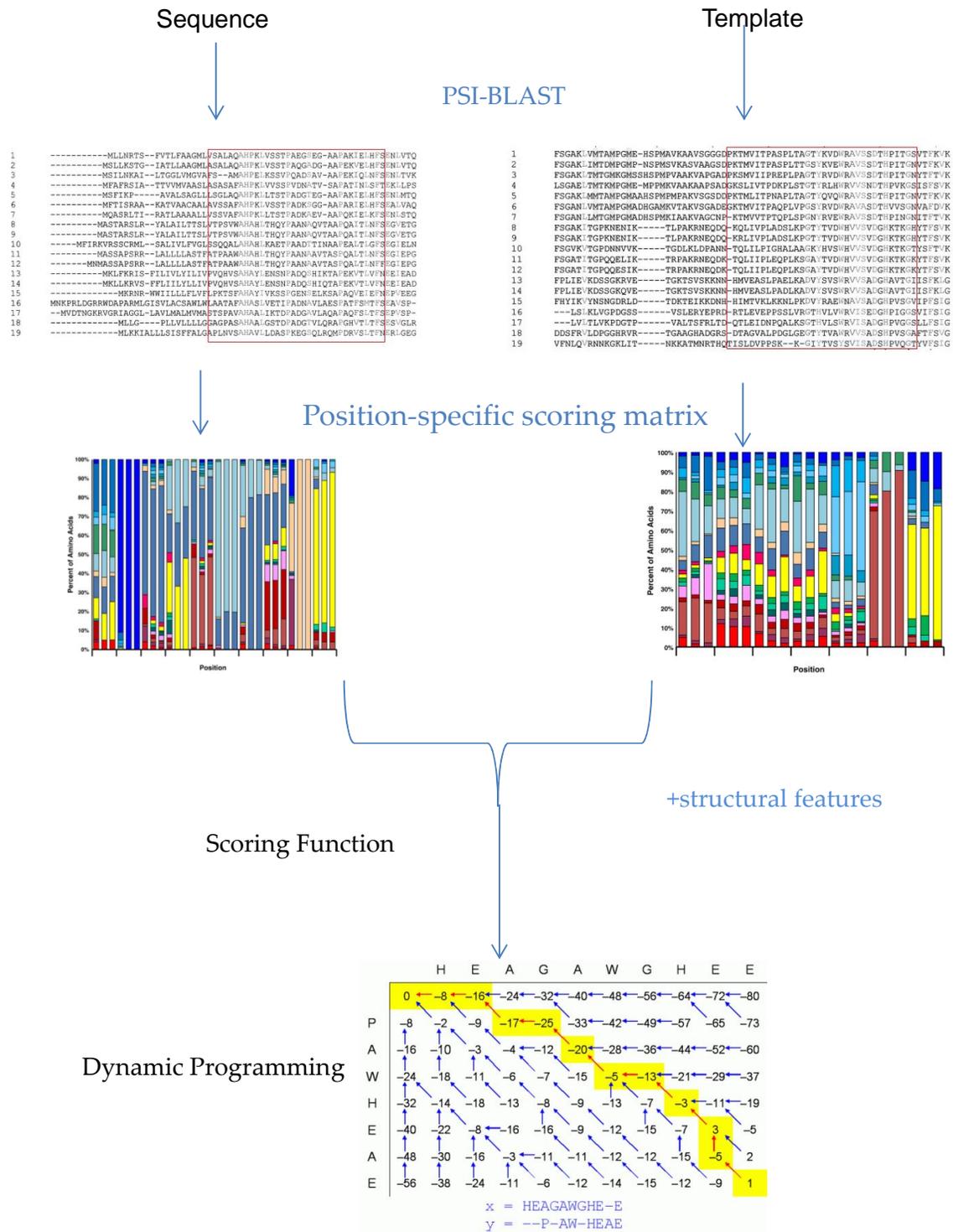

**Figure 1.5.** The standard flowchart for protein threading.

# Chapter 2

# Gradient tree boosting Conditional Random Fields for pairwise alignment

## 2.1 Introduction

The error of a template-based model comes from bad template selection and incorrect sequence-template alignment, in addition to the structural difference between the query protein and its template. At a high sequence identity (>50%), template-based models can be accurate enough to be useful for virtual ligand screening (Bjelic and Aqvist 2004; Caffrey, Placha et al. 2005), designing site-directed mutagenesis experiments (Skowronek, Kosinski et al. 2006), small ligand docking prediction, and function prediction (Skolnick, Fetrow et al. 2000; Baker and Sali 2001). When sequence identity is lower than 30%, it is difficult to recognize the best templates and generate accurate sequence-template alignments, so the resultant 3D models have a wide range of accuracies (Sanchez, Pieper et al. 2000; Chakravarty, Godbole et al. 2008). Pieper et al. have shown that 76% of all the models in MODBASE are from alignments in which the sequence and template share <30% sequence identity (Pieper, Webb et al. 2011). Therefore, to greatly enlarge the pool of useful models, it is essential to improve fold recognition and alignment methods for the sequence and template with <30% sequence identity. Considering that currently there are millions of proteins without experimental structures, even a slight improvement in prediction accuracy can have a significant impact on the large-scale structure prediction and related applications. As reported in (Melo and Sali 2007), even 1% improvement in the accuracy of fold assessment for the ~4.2 million models in MODBASE can correctly identify ~42 000 more models.

Various structural and evolutionary features have been widely used for template-based modeling to improve the alignment accuracy.

Most threading programs use a simple linear combination of these features as the scoring function, because the linear scoring function can be easily tuned and also can be efficiently optimized using dynamic programming algorithms. However, a linear scoring function cannot accurately account for the interdependency among features, which would lead to the sub-optimal performance for protein threading. It has been observed that some sequence and structure features (e.g., secondary structures and solvent accessibility) are highly correlated. To model the dependency among features, the SVM-align method (Yu, Joachims et al. 2008) explicitly enumerates hundred-thousands of complex features, which leads to the same number of model parameters to be trained. A complex feature is a combination of some basic features, e.g., secondary structure, solvent accessibility and amino acid type. However, a threading method with such a large number of parameters is not amenable to training since 1) it needs a large number of training examples to fit these parameters; 2) it needs careful tuning to avoid overfitting; and 3) the training process is highly time-consuming. Using such a complicated model, it is also computationally intensive to find the best sequence-template alignment between a protein pair, which makes the method unsuitable for protein structure prediction at the genome-wide scale. Furthermore, not all the features are equally important and some unimportant features may introduce noise into the model. An effective and compact scoring function to better exploit information of multiple features is thus required to further advance protein template-based modeling.

This chapter presents a nonlinear scoring function for protein threading, which not only can model dependency among various sequence and structure features, but also can be optimized efficiently using a dynamic programming algorithm. We fulfill this by modeling the protein threading problem using a probabilistic graphical model Conditional Random Fields (CRFs) (Lafferty, McCallum et al. 2001) and training this model using the gradient tree boosting algorithm proposed in (Dietterich, Hao et al. 2008). The resultant threading scoring function consists of only dozens of regression trees, which are automatically constructed during model training process to capture the nonlinear dependency among sequence and structure features. Experimental results indicate that by modeling feature interactions using regression trees, we can effectively leverage weak biological signals and greatly improve alignment accuracy and fold recognition rate.

## 2.2 Conditional Graphical Models for pairwise alignment

### *2.2.1 Conditional Random Fields*

Conditional random fields (CRFs) are probabilistic graphical models that have been extensively used in modeling sequential data (Lafferty, McCallum et al. 2001). Recently, CRFs have also been used to model various computational biology problems such as protein secondary structure prediction (Lafferty, Zhu et al. 2004), protein conformation sampling (Zhao, Li et al. 2008) and protein sequence alignment (Do, Gross et al. 2006). Different from canonical classification methods which predict the label of a single sample, a CRF can take neighboring samples or the context into account and predict labels for a set of related samples jointly. For example, the linear chain CRF predicts the sequence of secondary structure annotations of input amino acid sequence by jointly modeling the preference of secondary structure of several sequential residues.

For sequential data, linear CRFs can be seen as the discriminative and undirected variants of hidden Markov models. Consider the problem of learning to assign labels to a set of observation sequence. A HMM define a generative model for the joint probability distribution P(X,Y) where X and Y are the observation sequence and the label sequence respectively. In contrast, CRFs directly model the probability distribution P(Y|X) of labels conditioned on the observations. In this way, CRFs are more flexible than HMMs, allowing the relaxation of strong independence assumptions made by HMMs.

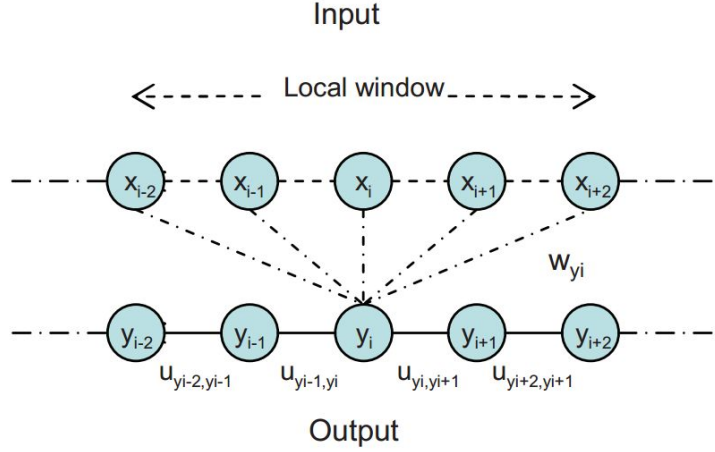

**Figure 2.1.** A diagram of a conditional random field model for sequential data.

***Graph structure, potential functions and conditional probability distribution.*** A CRF is associated with an undirected graph G=(V,E) in which each node corresponds to a random variable representing an element of labels Y. The structure of graph G can be arbitrary, provided that the conditional independencies of the label variables are modeled. To model the conditional distribution P(Y|X), a CRF factorizes the joint distribution of Y into a product of potential functions according to the structure of graph G. Each potential function measures the local preference of a subset of the label random variables. Formally, a linear CRF defines the probability of a particular label sequence Y given observation sequence X to be a normalized product of potential functions:

$$P(Y|X) = \frac{1}{Z(X)} \exp(\sum_{t=1}^{N} (\psi(Y,X,t) + \phi(Y,X,t))) \qquad (1)$$

$$\phi(Y,X,t) = \sum_{y} w_y^T f_y(Y,X,t) \qquad (2)$$

$$\psi(Y,X,t) = \sum_{y,y'} u_{y,y'} f_{y,y'}(Y,X,t) \qquad (3)$$

where N is the length of the sequence, $f_y(Y,X,t)$ and $f_{y,y'}(Y,X,t)$ are the feature functions defined on position $t$. For example, in Figure 2.1, $f_y(Y,X,t)$ is a feature defined on the neighborhood around position $t$; $f_{y,y'}(Y,X,t) = \delta[y_t = y]\delta[y_{t-1} = y']$ is the indicator function that describes the dependency between neighboring output labels. *Z(X)* is the normalization factor to make sure the distribution is valid. $w, u$ are the weights that parameterize the distribution and need to be determined in the training process.

***Inference and learning algorithms.*** Given the parameters and input X of a linear CRF model, the output labels Y can be easily inferred by Viterbi algorithm, which is a dynamic programming algorithm. Given a set of training data $T = \{(X, Y)\}$ of paired input and output sequences, the parameter of a linear CRF model can be optimized by maximizing the likelihood of the training data, which is also called the maximum likelihood principle. Gradient descent algorithms, such as L-BFGS, conjugate gradient descent and stochastic gradient descent (Sutton and McCallum 2012), have been widely used for this optimization. For linear CRF models, gradient can also be calculated efficiently by a dynamic programming algorithm called forward-backward algorithm (Sutton and McCallum 2012). More detailed discussions of the technical implementations can be found in a good review article (Sutton and McCallum 2012).

### 2.2.2 Modeling pairwise alignment

Although sequence alignments can be seen as a type of sequential data, linear CRFs cannot be directly used for protein alignment. Here we describe how to model the protein threading problem using conditional random fields. For a given pair of target and template, their sequence and structure features are called observations and their alignment is viewed as a sequence of labels.

Let s denote the target protein and its associated features, e.g., PSI-BLAST sequence profile, PSIPRED-predicted secondary structure (McGuffin, Bryson et al. 2000) and predicted solvent accessibility. Let t denote the template and its associated information, e.g., position-specific scoring matrix, solvent accessibility and secondary structure. Let $X = \{M, I_s, I_t\}$ be a set of three possible alignment states. Meanwhile, M indicates that two positions are matched and the two Is indicate insertion/deletion states. Is and It indicate insertions at sequence and template, respectively. We also tried to differentiate gaps at the two ends from gaps in the middle region by using four more states, but the resultant 7-state model is only slightly better than the three-state model. Let $a = \{a_1, a_2, \dots, a_L\}$ ($a_i \in X$) denote an alignment between s and t where $a_i \in X$ represents the state (or the label) at position i; for i-th alignment state, we use *is(i)* and *it(i)* to denote the corresponding residues on sequence and template respectively. For example, In Figure 2.2, *is(3)=1* and *it(3)=2* for the 3rd state Is (insertion at sequence). Our CRF model for threading defines the conditional probability of a given s and t as follows,

$$p(a \mid s,t) = \exp(\sum_i F(a_{i-1} \to a_i \mid s_{is(i)}, t_{it(i)}))/Z(s,t) \qquad (4)$$

where Z(s,t) is a normalization factor summing over all the possible alignments between s and t; F is the (un-normalized) log-likelihood of state transition in the alignment given the corresponding features. Traditionally, $F(a_{i-1} \to a_i \mid s_{is(i)}, t_{it(i)})$ is defined as a weighted sum of CRF features, which model the local state preference at alignment position i and the dependency of the state transition (from i-1 to i) on the target and template information of residues *is(i)* and *it(i)*. The features for function F are extracted from the corresponding residues and their neighbors. Once the model parameters are determined, we can find the best sequence-template alignment by maximizing $p(a \mid s,t)$, which can be done using a dynamic programming algorithm since $p(a \mid s,t)$ only models state transition between two adjacent positions.

## 2.3 A nonlinear scoring function for sequence alignment

Instead of explicitly enumerating thousands of complex features, we implicitly construct only a small number of important features using regression trees. The fitness between query residues and template residues is then determined through a cascade of decisions, thus more expressive than traditional scoring functions that can only combine features linearly. Then $F(a_{i-1} \rightarrow a_i \mid s_{is(i)}, t_{it(i)})$ is represented as a combination of regression trees instead of a linear combination of features. Each regression tree models the complex interactions among the basic features and each path from the tree root to a leaf corresponds to a single complex feature. See Figure 2.2 for an example of a regression tree for protein threading. This regression tree first checks whether the secondary structures of query and template residues are the same. Based on the decision, a children node then checks the evolutionary fitness between two residues according to a threshold and determines the final score denoting the likelihood whether they can be aligned together.

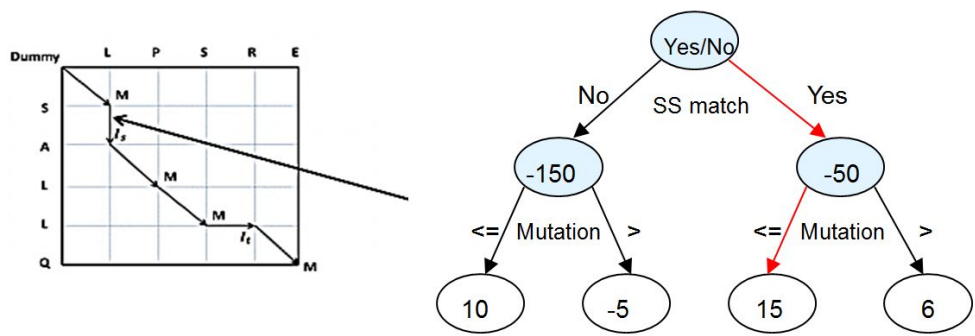

**Figure 2.2.** An example of regression tree for protein threading. The red arrows represent a decision path which evaluates the alignment fitness of two residues by secondary structure match and mutation scores.

We can build these regression trees automatically during the CRF training process using the gradient boosting algorithm proposed by (Dietterich, Hao et al. 2008). Only those important features emerge as a path (from tree root to a leaf) in the trees. The resulting scoring function has the form of a linear combination of regression trees. One advantage of this gradient tree boosting approach is that it is unnecessary to explicitly enumerate all the combinations of features. The important features can be automatically learned during the training process. In contrast, explicit enumeration may not generate features as good as those learned by regression trees. Another advantage is that we may avoid overfitting because of the ensemble effect of combining multiple regression trees and much fewer complex features used. The complexity of our method can also be controlled by limiting the total number of leaves or depth of trees. In this work, we set the depth up to 8. Finally, once the regression tree-based threading model is trained, we can find the best alignment very efficiently using the dynamic programming technique since there are only dozens of regression trees in the scoring function trained by our method.

To use the gradient tree boosting algorithm (Dietterich, Hao et al. 2008), we have $F(a_{i-1} \to a_i | s_{is(i)}, t_{it(i)})$ be a nonlinear function that calculates the log-likelihood of the ith alignment state given the target and the template information. In this new representation, there are no concepts of edge and label features. Instead, $F(a_{i-1} \to a_i | s_{is(i)}, t_{it(i)})$ is a linear combination of regression trees. To train such a model, we need to calculate the functional gradient of the conditional probability with respect to $F$. Using a similar technique described in (Dietterich, Hao et al. 2008), we have the following result.

**Lemma 2.1.** *Let u and v denote the alignment states at positions i-1 and i, respectively. The functional gradient of log-likelihood with respect to $F(u \to u | s,t)$ is given by*

$$\frac{\partial \ln p(a | s,t)}{\partial F(a_{i-1} = u \to a_i = v | s_{is(i)}, t_{it(i)})} = I(a_{i-1} = u, a_i = v) - P(a_{i-1} = u, a_i = v | s,t)$$

*where $I(a_{i-1} = u, a_i = v)$ is a 0-1 function. Its value equals to 1 if and only if in the training alignment the state transition from i-1 to i is $u \to v$. $I(a_{i-1} = u, a_i = v)$ is the predicted probability of the state transition $u \to v$ under current threading model.*

The functional gradient in Lemma 1 is easy to interpret. Given a training alignment, if the transition $u \to v$ is observed at position i, then ideally the predicted probability $P(a_{i-1}=u, a_i=v|s,t)$ should be 1 in order to make the functional gradient be 0 and thus, to maximize $p(a|s,t)$. Similarly, if the transition is not observed, then the predicted probability should be 0 to maximize $p(a|s,t)$. Given an initial $F(a_{i-1}=u \to a_i=v|s_{is(i)}, t_{it(i)})$, to maximize $p(a|s,t)$, we need to move F along the gradient direction. Since F is a function taking as input the sequence and structure features used for each alignment state, the gradient direction is also a function with the same input variables. We can use a regression tree T to fit the functional gradient with the corresponding input values being the sequence and template features around positions i − 1 and i. Then F is updated by F + wT where w is the step size and T is the gradient direction. The gradient tree boosting algorithm simply involves fitting regression trees to the difference between the observed and the predicted probabilities of each possible state transition. There are many possible functions that can fit a given set of data. Regression trees are chosen also because they are easy to interpret and can be quickly trained from a large number of examples. In addition, we can also control the tree depth or the number of leaves to avoid overfitting. Given a threading model and a training alignment, we can calculate $P(a_{i-1}=u, a_i=v|s,t)$ using the following forward-backward method. Let $\alpha(v, i)$ and $\beta(v, i)$ denote the probabilities of reaching state v at position i, starting from the N-terminal and C-terminal of the alignment, respectively. Both $\alpha(v, i)$ and $\beta(v, i)$ can be recursively calculated by standard dynamic programming algorithm. Given a set of training alignments, the gradient tree boosting algorithm to train the threading model is shown in Algorithm 1. The main component of the algorithm is to generate a set of examples to train a regression tree T for any feasible state transition $u \to v$. At any two adjacent positions of a training alignment, we generate an example by calculating $I(a_{i-1}=u, a_i=v) - P(a_{i-1}=u, a_i=v|s,t)$ as the response value and extracting sequence and structure features at the corresponding residues as the input values. Then we fit a regression tree to these examples and update F accordingly. The dynamic programming calculations are described in the following equations.

$$\alpha(v,1) = \exp(F(\emptyset \to v|s_1,t_1)) \tag{5}$$

$$\alpha(v,i) = \sum_u \exp(F(u \to v|s_{is(i)},t_{it(i)}))\alpha(u,i-1) \tag{6}$$

$$\beta(v,N) = 1 \tag{7}$$

$$\beta(v,i) = \sum_u \exp(F(v \to u|s_{is(i+1)},t_{it(i+1)}))\beta(u,i+1) \tag{8}$$

$$P(a_{i-1}=u, a_i=v \mid s,t) = \frac{\Gamma(u,i-1)\exp(F(u \to v \mid s_{is(i)},t_{it(i)}))\varsigma(v,i)}{Z(s,t)} \tag{9}$$

$$Z(s,t) = \sum_u \Gamma(u,0)\varsigma(u,0) \tag{10}$$

There are some tricky issues in building the regression trees due to the extremely unbalanced number of positive and negative examples. A training example is positive if its response value is positive or it is associated with an aligned position, otherwise negative. Given a training alignment 200 residues in each protein and 150 aligned positions, the ratio between the number of positive examples and that of negative ones is approximately (200+200−150)/(200×200×3)=0.002. This will result in serious bias in regression tree training. We employed two strategies to resolve this issue. One is to add more weights to the positive examples and the other is that we randomly sample a small subset of negative examples. Unlike the traditional CRF using L2 norm to regularize the model complexity and avoid overfitting, the complexity of our model is regularized by the tree depth. In building each regression tree, we use an internal 5-fold cross-validation procedure to determine the best tree depth. In our training process, the average tree depth is 4. Using such regularization, we can avoid overfitting in training the model.

```
function TreeBoostThreader(Data) // $Data = \{(a^j, s^j, t^j)\}$ where $j$ indicates the $j^{th}$ align-
ment example. $L^j$ is the length of the $j^{th}$ alignment.
for all state transition $u \to v$ do
    initialize $F_0^v(u,.,.) = 0$ //$u \to v$ is the feasible state transition at two adjacent positions
    //the second variable of $F_0^v(u,.,.)$ is the features extracted from the target protein
    //the third variable of $F_0^v(u,.,.)$ is the structure features from the template protein
end for
//training at most $M$ iterations
for $m$ from 1 to $M$ do
    for all state transition $u \to v$ do
        $S(u,v)$:=GenerateExamples($u$, $v$, Data)
        $T_m(u,v)$:=FitRegressionTree($S(u,v)$)
        $F_m^v(u,.,.) := F_{m-1}^v(u,.,.) + T_m(u,v)$
    end for
end for
return $F_M^v(u,s,t)$ as $F^v(u,.,.)$ for all $u \to v$
end function

function GenerateExamples(u,v,Data)
for all training alignments do
    for all $i$ from 1 to $L^j$ do
        Calculate $\alpha$ and $\beta$ according to Equations 6 and 8
        for all state transition $u \to v$ do
            Calculate $P(u,v|s,t)$ using Equation 9
            $\delta(i,u,v,s,t) = I(a_{i-1}=u, a_i=v) - P(a_{i-1}=u, a_i=v|s,t)$
            //$\delta$ is the response value to be fitted by a regression tree
            //$s(i)$ and $t(i)$ are the sequence and structure features around position $i$
            insert an example data entry $(s(i), t(i), \delta)$ into $S(u,v)$
        end for
    end for
end for
return $S(u,v)$
end function
```

**Algorithm 2.1.** Gradient tree boosting training algorithm

## 2.4 Evolutionary and structural features

We use both evolutionary information and structure information to build regression trees for our threading model. We generate sequence profiles as follows. We run PSI-BLAST with five iterations and E-value 0.001 to generate position specific score matrix (PSSM) for a template and position specific frequency matrix (PSFM) for a target. PSSM(i, a) is the mutation potential for amino acid a at template position i and PSFM(j, b) is the occurring frequency of amino acid b at target position j. The secondary structure and solvent accessibility of a template is calculated by the DSSP program (Kabsch and Sander 1983). For a target protein, we use PSIPRED (McGuffin, Bryson et al. 2000) and SSpro (Pollastri, Baldi et al. 2002) to predict its secondary structure and solvent accessibility, respectively.

### 2.4.1 Features for match state

We use the following features to build regression trees for a state transition to a match state. Suppose that template position i is aligned to target position j.

1. *Sequence profile similarity.* The sequence profile similarity score between two positions is calculated by $\Sigma_a$ PSSM(i, a) × PSFM(j, a).

2. *Contact capacity score.* The contact capacity potential describes the hydrophobic contribution of free energy, measured by the capability of a residue make a certain number of contacts with other residues in a protein. The two residues are in physical contact if the spatial distance between their C$\beta$ atoms is smaller than 8 Angstrom. Let CC(a, k) denote the contact potential of amino acid a having k contacts (Section 3 in (Xu 2005)). The contact capacity score is calculated by $\Sigma_a$ CC(a, c) × PSFM(j, a) where c is the number of contacts at template position i.

3. *Environmental fitness score.* This score measures how well we can align one target residue to a template local environment, which is defined by a combination of three secondary structure types and three solvent accessibility states. Let F(env, a) denote the environment fitness potential for amino acid a being in a local environment env (see Section 3 in (Xu 2005)). The environment fitness score is given by $\Sigma_a$ F(env$_i$, a) × PSFM(j, a).

4. *Secondary structure match score.* Supposing the secondary structure type at template position i is ss, then the predicted likelihood of ss at target position j is used as the secondary structure match score.

5. *Solvent accessibility match score.* This is a binary feature used to indicate if the template position and the target position are in the same solvent accessibility state.

*2.4.2 Features for gap state*

The simplest scoring model for gap penalty is an affine function o + e×g where o is the gap open penalty, e gap extension penalty and g the number of gapped positions. To improve alignment accuracy, some threading programs, such as SALIGN (Marti-Renom, Madhusudhan et al. 2004), use a context-specific gap penalty function while others such as HHpred (Soding 2005), SP5 (Zhang, Liu et al. 2008) and the new PROSPECT (Ellrott, Guo et al. 2007) use a position-specific gap penalty model. In our threading model, we use a more sophisticated context-specific gap penalty function. The regression trees for a state transition to an insertion state at the template depend on the following features on the template side: secondary structure type, solvent accessibility, amino acid identity and hydropathy counts (Do, Gross et al. 2006). Similarly, the regression trees for a state transition to an insertion state at the target depend on the following features on the target side: predicted secondary structure likelihood scores, predicted solvent accessibility, amino acid identity and hydropathy counts.

## 2.5 Experiments

*2.5.1 Model training*

Similar to CONTRAlign (Do, Gross et al. 2006), our boosting-based threading model does not need a large data set for training. The alignment accuracy on the validation set does not increase with respect to the training set size as long as it is at least 30. We arbitrarily choose 30 protein pairs from the PDB as our training set and 40 pairs as the validation set. The average size of a protein contains 200 residues. In the training set, 20 pairs are similar at the same fold level but different superfamily level according to the SCOP classification (Murzin, Brenner et al. 1995). The other 10 pairs are similar at the same superfamily but different family level. Any two proteins in the training and validation set have sequence identity less than 30%. Reference alignments are built by the structural alignment program TM-align (Zhang and Skolnick 2005). We also guarantee that the proteins used for model training have no high sequence identity (30%) with the proteins in the Prosup (Lackner, Koppensteiner et al. 2000) and SALIGN (Marti-Renom, Madhusudhan et al. 2004) benchmarks.

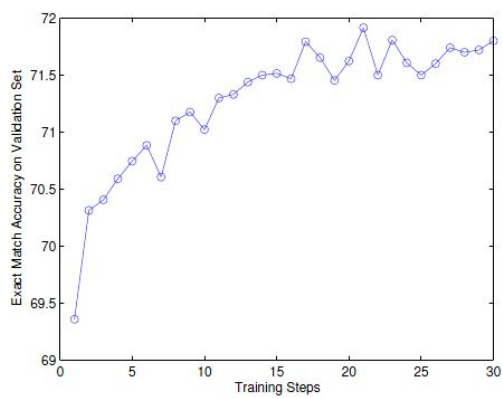

**Figure 2.3**. The alignment accuracy of the models on the validation data set during training.

As shown in Figure 2.3, the training process runs very fast. It takes approximately two minutes per iteration and achieves very good alignment accuracy after only six or seven iterations. The alignment accuracy reaches the best value after 21 iterations. More training iterations do not improve alignment accuracy but result in more regression trees. The more regression trees used in the threading model, the more running time will be needed to align a protein pair. As a result, we choose the model trained after 21 iterations as our final threading model. For each state transition, the model has twenty-one regression trees with an average depth four.

*2.5.2 Performance*

To compare our method with other state-of-art threading programs, we evaluate them on two popular benchmarks: Prosup (Lackner, Koppensteiner et al. 2000) and SALIGN (Marti-Renom, Madhusudhan et al. 2004). The Prosup benchmark has 127 protein pairs with structural alignments generated by Prosup. The SALIGN benchmark contains 200 protein pairs. On average, two proteins in a pair share 20% sequence identity and 65% of structurally equivalent $C\alpha$ atoms superposed with RMSD 3.5A. We used TM-align to generate structural alignments for the SALIGN benchmark. The SALIGN benchmark is more difficult than the Prosup benchmark because it includes many pairs of proteins with very different sizes. To evaluate the alignment quality, we use the exact match accuracy which is computed as the percentage of one-to-one match positions in the reference alignments. We also evaluate the 4-offset match accuracy, which is defined as the percentage of the matches within 4 positions shift from one-to-one match. Table 2.1 compares the performance of various alignment methods on the Prosup benchmark. Our method, denoted as BoostThreader, shows a significant improvement over the others. The absolute improvement over SP3/SP5, a leading threading program, is more than 5%. The major difference between our method and SP3/SP5 is that SP3/SP5 linearly combines various sequence and structure features as its scoring function while our method uses a nonlinear scoring function. CONTRAlign (Do, Gross et al. 2006) is run locally with the default hydropathy model. CONTRAlign mainly aims at sequence alignment, so it is not surprising that its performance is not as competitive as some leading threading methods. The results of other methods are taken from (Qiu and Elber 2006; Liu, Zhang et al. 2007; Zhang, Liu et al. 2008). Also as shown in the right three columns of Table 1, our method also has the best alignment accuracy on the SALIGN benchmark. This benchmark contains many pairs of proteins with very different sizes, which is the major reason why RAPTOR (Xu, Li et al. 2003) performs badly on this benchmark.

**Table 2.1**. Alignment accuracy (%) of our method BoostThreader and other alignment methods on the Prosup and SALIGN benchmarks.

| | Prosup | | | SALIGN | | |
|---|---|---|---|---|---|---|
| Methods | Exact | 4-offset | Methods | Exact | 4-offset |
| CONTRAlign | 52.79 | 69.42 | CONTRAlign | 44.38 | 57.37 |
| PSIBLAST | 35.60 | | PSIBLAST | 26.10 | |
| SPARKS | 57.20 | | SPARKS | 53.10 | |
| SSALGN | 58.30 | | SALIGN | 56.40 | |
| RAPTOR | 61.30 | 79.32 | RAPTOR | 40.20 | 59.80 |
| SP3 | 65.30 | 82.20 | SP3 | 56.30 | 56.60 |
| SP5 | 68.70 | | SP5 | 59.70 | |
| BoostThreader | **74.05** | **88.90** | BoostThreader | **63.60** | **79.01** |

*Fold recognition.* We also evaluate the fold recognition rate of our new method BoostThreader on the Lindahl's benchmark (Lindahl and Elofsson 2000), which contains 976 proteins. Any two proteins in this set share less than 40% sequence identity. All-against-all threading of these proteins can generate 976 × 975 pairs. After generating the alignments of all the pairs using BoostThreader, we rank all the templates for each sequence using a similar to (Xu 2005) and then evaluate the fold recognition rate of our method. When evaluating the performance in the superfamily level, all the templates similar at the family level are ignored. Similarly, when we evaluate the performance at the fold level, all the templates similar in the superfamily or family level are ignored. "Top 1" means that the only the first-ranked templates are evaluated while "Top 5" indicates that the best templates out of the top 5 are evaluated. As shown in Table 2, our method performs well at all three similarity levels. The fold recognition rate of our new method is much better than SP3/SP5, HHpred and RAPTOR, especially at the superfamily and fold levels. These three programs performed very well in recent CASPs.

**Table 2.2.** Fold recognition rate (%) of various threading programs. The PSI-BLAST, SPARKS, SP3, SP5 and HHpred results are taken from (Zhang, Liu et al. 2008). The FOLDpro, HMMER, FUGUE, SAM-98 results are from (Cheng and Baldi 2006). The RAPTOR and PROSPECT-II results are from (Xu 2005).

| Methods | Family | | Superfamily | | Fold | |
|---|---|---|---|---|---|---|
| | Top1 | Top5 | Top1 | Top5 | Top1 | Top5 |
| PSIBLAST | 71.2 | 72.3 | 27.4 | 27.9 | 4.0 | 4.7 |
| HMMER | 67.7 | 73.5 | 20.7 | 31.3 | 4.4 | 14.6 |
| SAM-98 | 70.1 | 75.4 | 28.3 | 38.9 | 3.4 | 18.7 |
| THREADER | 49.2 | 58.9 | 10.8 | 24.7 | 14.6 | 37.7 |
| FUGUE | 82.2 | 85.8 | 41.9 | 53.2 | 12.5 | 26.8 |
| PROSPECT-II | 84.1 | 88.2 | 52.6 | 64.8 | 27.7 | 50.3 |
| SPARKS | 81.6 | 88.1 | 52.5 | 69.1 | 24.3 | 47.7 |
| SP3 | 81.6 | 86.8 | 55.3 | 67.7 | 28.7 | 47.4 |
| FOLDpro | 85.0 | 89.9 | 55.5 | 70.0 | 26.5 | 48.3 |
| SP5 | 81.6 | 87.0 | 59.9 | 70.2 | 37.4 | 58.6 |
| HHpred | 82.9 | 87.1 | 58.8 | 70.0 | 25.2 | 39.4 |
| RAPTOR | **86.6** | 89.3 | 56.3 | 69.0 | 38.2 | **58.7** |
| BoostThreader | 86.5 | **90.5** | **66.1** | **76.4** | **42.6** | 57.4 |

# Chapter 3

# Low-homology protein threading

## 3.1 Introduction

As discussed in the previous chapter, accuracy of protein alignment is majorly determined by the scoring function used to drive sequence-template alignment. When the sequence and template are not close homologs, their alignment can be significantly improved by incorporating homologous information (i.e. sequence profile) into the scoring function. HHpred (Soding 2005), possibly the best profile-based method, uses only sequence profile and predicted secondary structure for remote homolog detection. It works very well when proteins under consideration have a large amount of homologous information in the public sequence databases, but not as well when proteins under consideration are low-homology. A protein is low-homology if there is no sufficient homologous information available for it in the sequence databases.

The capability of predicting low-homology proteins without close homologs in the PDB is particularly important because (i) a large portion of proteins in the PDB, which will be used as templates, are low-homology; and (ii) a majority number of the Pfam (Sammut, Finn et al. 2008) families without solved structures are low-homology (see Section 3.2). Therefore, to predict structure for proteins in Pfam using templates, it is essential to have a method that can work well on low-homology proteins. In addition, the class of low-homology proteins may represent a substantial portion of metagenomics sequences of microbes (e.g. Staphylococcus aureus) generated from numerous metagenomics projects. Because (i) its sequence profile does not contain enough evolutionary information to link it to remote homologs in the PDB; and (ii) its predicted secondary structure (and other predicted structural features) usually has low accuracy as the secondary structure (and other predicted structural features) is usually predicted from homologous information, such as sequence profile, it is very difficult to predict structure of low-homology proteins with a satisfactory accuracy. Many existing template-based modeling methods, including those mentioned in Chapter 1 (for instance MUSTER (Wu and Zhang 2008), Phyre2 (Kelley and Sternberg 2009) and SPARKS/SP3/SP5 (Zhang, Liu et al. 2004; Zhou and Zhou 2004; Zhou and Zhou 2005; Zhou and Zhou 2005; Zhang, Liu et al. 2008)), aim at going beyond pure profile-based methods by combining homologous information with a variety of structural information. However, recent CASP evaluations (Moult, Fidelis et al. 2005; Moult, Fidelis et al. 2007; Moult, Fidelis et al. 2009; Moult, Fidelis et al. 2011) demonstrate that HHpred actually is as good as if not better than these threading methods. Clearly, it is very challenging to outperform HHpred a lot by simply adding structural information into template-based methods. In fact, Ginalski et al. (Ginalski, Grishin et al. 2005) claimed that "presently, the advantage of including the structural information in the fitness function cannot be clearly proven in benchmarks".

To improve the prediction for these low-homology proteins, this chapter describes a novel profile-dependent scoring function for protein threading. This scoring function automatically determines the relative importance of structural information according to the amount of homologous information available. When proteins under consideration are low-homology, our method will rely more on structural information; otherwise, homologous information. It enables us to significantly advance template-based modeling over profile-based methods such as HHpred, especially for low-homology proteins. This method is mainly built by our previous approaches described in Chapter 2 and is also incorporated into the new BoostThreader program.

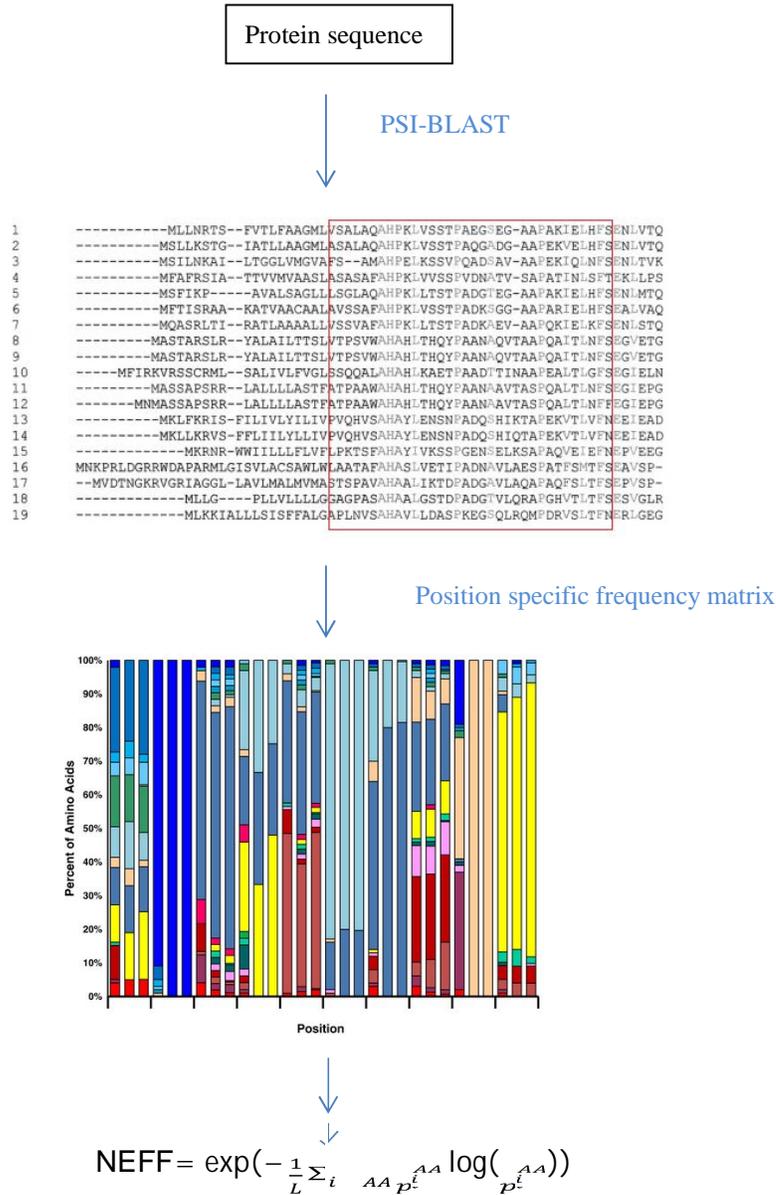

**Figure 3.1.** Calculation of NEFF. The protein sequence is firstly searched against the non-redundant sequence database. A multiple sequence alignment is generated by PSI-BLAST. A position specific frequency matrix is then calculated from the alignment. The NEF is calculated as the exponential of average entropy over all columns.

## 3.2 Number of EFFective homologs (NEFF)

NEFF is not a new concept. It has already been used by PSI-BLAST (Altschul, Madden et al. 1997) to measure the amount of homologous information available for a protein. The relationship between NEFF and the modeling capability of a profile-based method has also been studied before (Casbon and Saqi 2004; Sadreyev and Grishin 2004). NEFF can be interpreted as the effective number of non-redundant homologs of a given protein and be calculated from the multiple sequence alignment with the homologs. The homologs are detected in the NCBI non-redundant (NR) database by PSI-BLAST (five iterations and E-value 0.001). NEFF is calculated as the exponential of entropy averaged over all columns of the multiple sequence alignment, so in this sense NEFF can also be interpreted as the entropy of a sequence profile derived from the multiple sequence alignment. NEFF for a protein is a real value ranging from 1 to 20. A protein with a small NEFF value (<6) is low-homology since we cannot obtain sufficient homologous information for it from existing protein sequence databases.

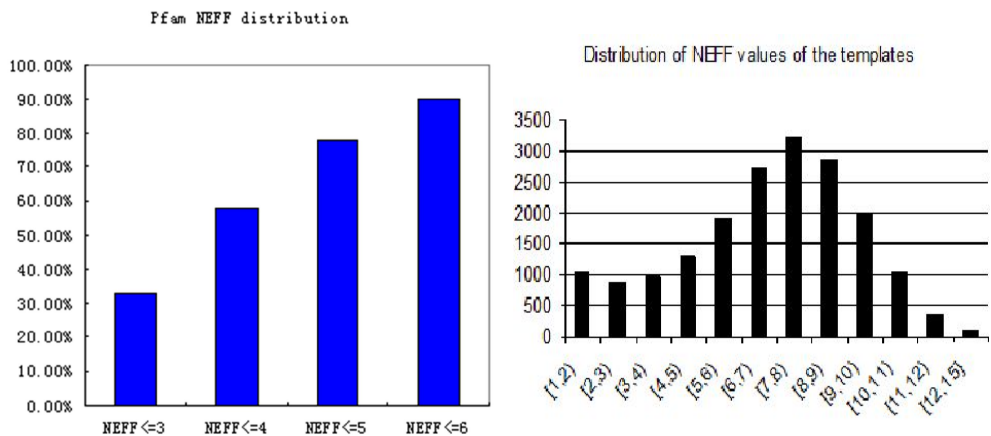

**Figure 3.2.** Distributions of NEFF in Pfam database and template database.

The Pfam (version: 23.0) contains 10 000 families covering 75% protein sequences in UniProt (Sammut, Finn et al. 2008). Among the 6600 Pfam families without solved structures, ~90, ~78, ~58 and ~33% of them have NEFF smaller than 6, 5, 4 and 3, respectively. Among the ~26000 HHpred templates (i.e. a set of representative structures in the PDB), ~36% of them have NEFF <6 (see Figure 3.2). There are also ~25% protein sequences in UnitProt not covered by the Pfam database. Many of these sequences are singletons (i.e. products of orphan genes) and thus, have NEFF=1. In the foreseeable future, many of the low-homology proteins or protein families (i.e. NEFF ≤6) will not have solved structures. Also the correlation between the hardness of homology can be seen from CASP evaluations. For example, Figure 3.3 shows that the correlation between the NEFF values and the hardness of the targets in CASP8 is significant, where the hardness is calculated as the model quality (GDT-TS score) of HHpred's top predictions. NEFF values of most hard targets in CASP8 are less than 6 while most easy targets have NEFF values greater than 6. Consequently, to elucidate the structures of these low-homology proteins (or protein families) and expand our knowledge of the overall protein structure space, it is a pressing need to develop a protein threading method that can work well on such proteins.

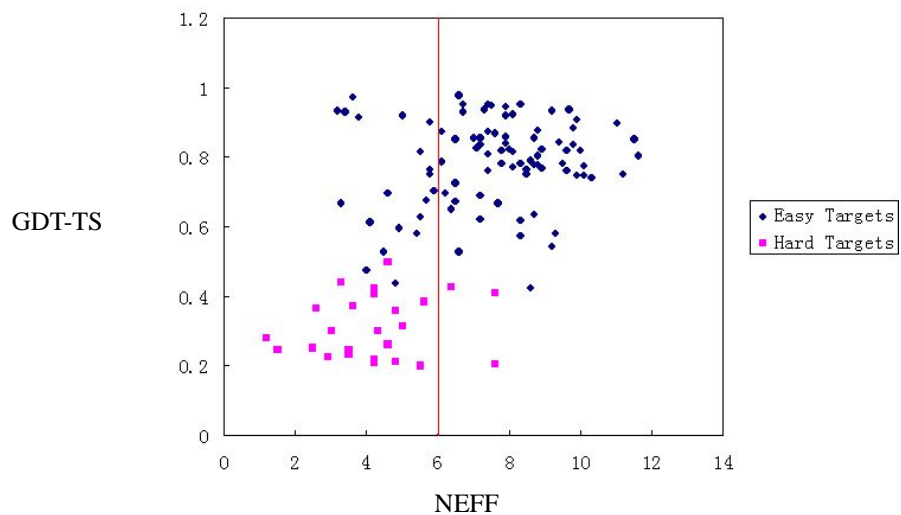

**Figure 3.3.** Correlation between NEFF and the prediction quality by HHpred on CASP8 targets. The magenta points represent the difficult targets and the blue points denote the easy targets. The classification of hard and easy targets is adopted from Yang Zhang's CASP8 assessment website.
(zhanglab.ccmb.med.umich.edu/casp8/).

## 3.3 Implementation of the protein threading model

We create a low-homology protein threading model in the way introduced in Chapter 2. We incorporate NEFF and a set of other new features into the CRF model. We train this model by maximizing the occurring probability of a set of reference alignments. Gradient tree boosting algorithm is applied to estimate the ensemble of regression trees as the threading scoring function.

### *3.3.1 Features for a match state*

In addition to the features (profile similarity, secondary-structure similarity, solvent accessibility similarity and environmental fitness score) described in previous Chapter (also described in (Peng and Xu 2009)), we use the following extra information to estimate the probability of one template position being aligned to one target position. In order to determine the relative importance of homologous and structure information, the NEFF values of both the sequence and template are used as features. When NEFF is large, our threading method will count more on homologous information, otherwise on structure information. We use the CC50 matrix developed by Kihara group (Tan, Huang et al. 2006) to calculate similarity between the sequence and template. This matrix is a statistical potential-based amino acid similarity matrix, originally designed for aligning distantly related protein sequences. One element $CC50[a][b]$ in this matrix is the similarity score between two amino acids $a$ and $b$, which is computed as the correlation coefficient of the pairwise contact potentials of these two amino acids. We also use a structure-based substitution matrix (Prlic, Domingues et al. 2000; Tan, Huang et al. 2006) to improve alignment accuracy when the sequence and template are distantly related. This scoring matrix is derived by a similar procedure as the BLOSUM matrices (Henikoff and Henikoff 1992) are done, based upon the structure alignments of structurally similar protein pairs. It is also shown to be more sensitive than BLOSUM in remote homolog detection.

### *3.3.2 Features for a gap state*

The gap event is related to multiple factors. Some studies have indicated that a gap event is related to its local sequence and structure context. For example, SSALN (Qiu and Elber 2006) uses a context-specific gap penalty model, in which a gap event depends on secondary structure and solvent accessibility. Other methods, such as HHpred and the method described by Ellrott *et al*. (Ellrott, Guo et al. 2007), use a position-specific gap penalty model, which contains evolutionary information of a protein. In the previous section or (Peng and Xu 2009), only context-specific gap penalty is used. In this work, we use both context-specific and position-specific gap penalty and then use the NEFF to determine their relative importance. If the NEFF is large, we will rely more on position-specific gap penalty (i.e. homologous information); otherwise, context-specific gap penalty (i.e. structure information). To calculate the position-specific gap penalty of a protein, we run PSI-BLAST with the query protein (with five iterations and *E*-value 0.001) against the NCBI non-redundant database and generate a multiple sequence alignment. Then we calculate the probability of a gap event at each residue as the ratio between the number of the gap events and the number of sequences in the multiple sequence alignment. For context-specific gap penalty, we estimate the occurring probability of an insertion at the template using secondary-structure type, solvent accessibility, amino acid identity and the count of neighboring hydrophilic residues (Do, Gross et al. 2006). In addition, we use a binary value to indicate if a residue is in the structurally compact core region or not. A core residue is usually more conserved and shall be. Similarly, we estimate the occurring probability of an insertion at the target using predicted secondary, predicted solvent accessibility, amino acid identity and the count of neighboring hydrophilic residues.

### 3.3.3 Geometric constraints

When the sequence and template are not close homologs, their alignment usually contains displaced gap opening or ending positions. Even a single displaced gap in an alignment may result in a big decrease of the quality of the predicted 3D model. The template provides some geometric information that can be used to improve alignment accuracy. Suppose that two adjacent sequence positions are aligned to two template positions $j_1$ and $j_2$ ($j_2>j_1+1$), respectively. Since the distance between two adjacent $C\alpha$ atoms is around 3.8 Angstrom, the two $C\alpha$ atoms at $j_1$ and $j_2$ should not be far apart. To tolerate some alignment errors, we use 7 Angstrom (instead of 3.8 Angstrom) as the distance threshold for such two $C\alpha$ atoms. We enforce this physical constraint when generating the optimal alignment between the sequence and template by the dynamic programming algorithm. All the alignments violating this physical constraint are discarded. Our experiments indicate that by applying this constraint, we can greatly improve alignment accuracy for some threading instances.

*3.3.4 Template selection*

After aligning the target to all templates in a database constructed from PDB with 95% sequence identity cut-off, we need to choose the best alignment, from which we can build a 3D model for the target. We use a neural network model to predict the quality, measured by TM-score (Zhang and Skolnick 2005), of the 3D model built by MODELLER from our sequence-template alignment and then use the predicted quality to rank all the alignments for the given target. We predict the TM-score using the following alignment-dependent features: sequence identity, distribution of various per-position scores such as mutation score, solvent accessibility score, secondary-structure similarity score and distribution of gap sizes. In addition, we feed the NEFF values of both the target and the template into our neural network, in order to determine the relative importance of homologous and structural information. We trained our template selection method using the data set generated by RAPTOR (Xu, Li et al. 2003) for both CASP6 and CASP7 targets. Tested on these targets (using cross-validation), the absolute prediction error of TM-score is ~0.045 on average (data not shown). The correlation coefficient between the predicted TM-score and the real one is above 0.9 on all alignments and 0.8 on low-quality ones.

## 3.4 Experiments

### 3.4.1 Training and validation datasets

We choose 66 protein pairs from the PDB as the training set and 50 pairs as the validation set. The NEFF (i.e. the diversity of sequence profiles) values of these 66 pairs of proteins are distributed uniformly between 1 and 11. This is very important in order to avoid structural information being dominated by homologous information. In the training set, 46 pairs are in the same fold but different superfamily level by the SCOP classification (Murzin, Brenner et al. 1995). The other 20 pairs are in the same superfamily but different family level. Any two proteins in the training and validation set have sequence identity <30%. The proteins used for model training and validation have no high sequence identity (<30%) with the proteins in the Prosup (Lackner, Koppensteiner et al. 2000) and SALIGN (Marti-Renom, Madhusudhan et al. 2004) benchmarks and the CASP8 targets. We use TM-align (Zhang and Skolnick 2005) to build a reference alignment for a protein pair in SALIGN.

### 3.4.2 Performance on two public benchmarks

We first test our method on two public benchmarks: Prosup (Lackner, Koppensteiner et al. 2000) and SALIGN (Marti-Renom, Madhusudhan et al. 2004) that are also used in original BoostThreader evaluation in Chapter 2. We evaluate our new BoostThreader using both reference-dependent and reference-independent alignment accuracy. The reference-dependent alignment accuracy is calculated as the percentage of correctly aligned positions judged by reference alignments, which are generated by structural alignment programs. To evaluate the reference-independent alignment accuracy, we first build a 3D model for the sequence in a protein pair using MODELLER (Sali 1995) from its alignment to the template and then evaluate the quality of the resultant 3D model using TM-score (Zhang and Skolnick 2005) and GDT-TS score (Zemla, Venclovas et al. 1999). Since our ultimate goal is to predict the 3D structure for a target protein, reference-independent alignment accuracy is a better measurement than reference-dependent alignment accuracy.

**Table 3.1.** Reference-dependent alignment accuracy comparisons on two public benchmarks.

| ProSup | | SALIGN | |
|---|---|---|---|
| Methods | Acc | Methods | Acc |
| PSIBLAST | 35.60 | PSIBLAST | 26.10 |
| ContraAlign | 52.79 | ContraAlign | 44.38 |
| SPARKS | 57.20 | SPARKS | 53.10 |
| SSALGN | 58.30 | SALIGN | 56.40 |
| RAPTOR | 61.30 | RAPTOR | 40.20 |
| SP3 | 65.30 | SP3 | 56.30 |
| SP5 | 68.70 | SP5 | 59.70 |
| HHpred | 69.04 | HHpred | 62.98 |
| **Our work** | **76.08** | **Our work** | **64.40** |

As shown in Table 3.1, our method shows a significant advantage over the other methods. The absolute improvement over our own RAPTOR threading program (Xu, Li et al. 2003) is at least 24%. Our method is also better than the CASP-winning methods SP3 and SP5 by 16.5% (14.4%) and 10.7% (7.9%) on ProSup (SALIGN), respectively. The results of SPARKS/SP3/SP5 are taken from Zhang *et al.* (2008). We also compare reference-independent alignment accuracy which evaluates the quality of a model by comparing it to the native structure and yields a number between 0 and 1. The higher the number, the better quality the model has. The models generated by our new method in total have TM-score 66.77 and TM-score 132.85 on Prosup and SALIGN, respectively. By contrast, HHpred achieves TM-score 56.44 and 119.83 on Prosup and SALIGN, respectively. Our method is better than HHpred by 18.3 and 10.9% on ProSup and SALIGN, respectively. A student's *t*-test indicates that our method excels HHpred with *P*-values being $3.77E-11$ and $9.83E-13$, respectively. To examine the performance of our method and HHpred with respect to the amount of homologous information, we divide the test protein pairs in the ProSup and SALIGN sets into 10 groups according to their NEFF values: [1,2), [2,3),…, [9,10), [10,20] (see Figure 3.4). The NEFF of a protein pair is defined as the minimum NEFF of the target and template. Out of the 327 test protein pairs, 15, 26, 53, 72 and 114 pairs have NEFF smaller than 2, 3, 4, 5 and 6, respectively. Then we calculate the average reference-independent alignment accuracy (measured by TM-score) of all the pairs in each group. As shown in Figure 3.2, when either the target or template has a small NEFF (<6), on average our method can generate much better 3D models than HHpred. When NEFF <2, the model quality of our method is almost 100% better than HHpred. When NEFF <3, the model quality of our method is at least 50% better than HHpred. Our method also performs as well as HHpred on high homology targets (i.e. NEFF >7). According to Skolnick group's study, a model with TM-score ~0.4 can be used for functional study while a model with TM-score ~0.2 is almost random. This implies that when NEFF <2, the HHpred models are almost random while our method can generate models useable for functional study. Since ~90% of the Pfam families without solved structures have NEFF <6, our method can improve over HHpred on most Pfam families. This study indicates that we can significantly advance the modeling capability of low-homology proteins with NEFF≤3, which represents approximately one-third of the Pfam families without solved structures.

### 3.4.3 Performance on CASP8 dataset

To further demonstrate the advantage of our method, we compare it with the top 14 CASP8 servers (see **Table 3.2**). Among these servers, only HHpred2, MUSTER and Phyre2 are pure threading-based methods. Other servers use a combination of multiple structure prediction techniques including consensus methods, multiple-template modeling, template-free modeling and model refinement. For example, Zhang-Server (Zhang 2008; Zhang 2009) first does a consensus analysis of the results generated by ~10 individual threading programs (Wu and Zhang 2007) and then refines models using distance restraints extracted from top templates by sampling-based optimization. Similar to Zhang-Server, the two TASSER programs (Zhou and Skolnick 2009) uses the results from two threading programs PROSPECTOR (Skolnick and Kihara 2001)) and SP3 (Zhou and Zhou 2005). Robetta (Raman, Vernon et al. 2009) first generates a template-based model using HHpred and then does model refinement. Robetta also runs template-free modeling if a reliable template cannot be detected. Phyre-de-novo combines the output of both HHpred and Phyre2 and in case no good template identified, also does template-free modeling. The three MULTICOM programs (Cheng 2008) (MUProt, MC-CLUSTER and MC-REFINE) use multiple threading programs, multiple-template techniques, model clustering and template-free modeling. Our RAPTOR++ (Xu, Peng et al. 2009) program uses three in-house threading programs and then employs multiple-template technique for easy targets and template-free modeling for very hard targets. TM-score is used to evaluate the prediction accuracy.

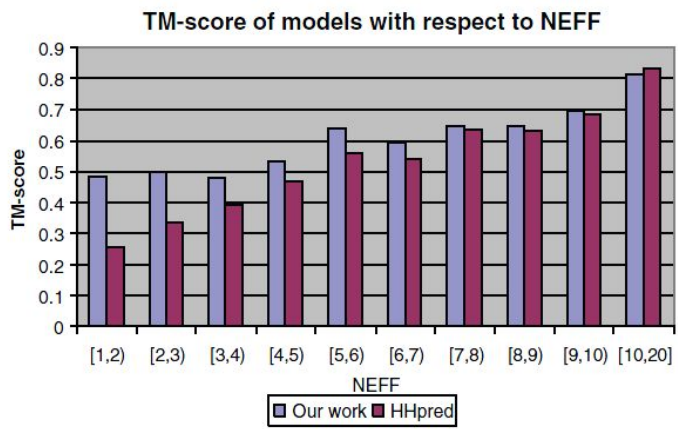

**Figure 3.4.** Reference-independent alignment accuracy (TM-score) comparison with HHpred. Each bin contains the protein paris with a specific NEFF range on Prosup and SALIGN datasets.

**Table 3.2.** Average TM-score of our method and the CASP8 top servers on 119 CASP8 targets with respect to NEFF

| NEFF | ≤2 | ≤3 | ≤4 | ≤5 | All |
|---|---|---|---|---|---|
| #targets | 2 | 6 | 16 | 33 | 119 |
| Zhang-Server | 0.243 | 0.278 | 0.505 | 0.501 | 0.711 |
| **Our work** | **0.291** | **0.336** | **0.521** | **0.486** | **0.694** |
| pro-sp3-TASSER | 0.248 | 0.247 | 0.471 | 0.476 | 0.691 |
| RAPTOR++ | 0.264 | 0.279 | 0.491 | 0.469 | 0.683 |
| METATASSER | 0.262 | 0.275 | 0.478 | 0.457 | 0.678 |
| ROBETTA | 0.270 | 0.262 | 0.489 | 0.470 | 0.676 |
| HHpred2 | 0.265 | 0.238 | 0.480 | 0.459 | 0.675 |
| Phyre-de-novo | 0.229 | 0.267 | 0.475 | 0.455 | 0.670 |
| MUSTER | 0.207 | 0.250 | 0.477 | 0.452 | 0.670 |
| MC-REFINE | 0.255 | 0.286 | 0.485 | 0.454 | 0.668 |
| HHpred5 | 0.275 | 0.225 | 0.475 | 0.446 | 0.668 |
| MC-CLUSTER | 0.212 | 0.286 | 0.489 | 0.455 | 0.667 |
| HHpred4 | 0.264 | 0.222 | 0.475 | 0.454 | 0.667 |
| MUProt | 0.254 | 0.271 | 0.478 | 0.454 | 0.664 |
| Phyre2 | 0.258 | 0.254 | 0.473 | 0.448 | 0.653 |

For fair comparisons, our new threading method used the NCBI NR and a template database generated before CASP8 started (i.e. May 2008). We evaluated the model quality of the 119 CASP8 targets using TM-score. The model quality of the CASP8 servers is downloaded from Zhang's CASP8 website (zhanglab.ccmb.med.umich.edu/casp8/). We exclude T0498 and T0499 from evaluation because they have been discussed in (Alexander, He et al. 2007) well before CASP8 started. By comparing our method with Zhang-Server, we can see how far away our new method is from the best server in the community, although it is unfair to compare our single-template-based method with a modeling method using multiple techniques. By comparing our method with the three mainly-threading-based methods HHpred2, MUSTER and Phyre2, we can see how much we have advanced the state-of-the-art of protein threading. This is important since all the top CASP8 servers including Zhang-Server heavily depend on single-template-based threading methods.

Further we investigate the performance on low-homology proteins. As shown in Tables 3.2, if only the low-homology targets (NEFF ≤4) are evaluated, our method outperforms all the top CASP8 servers including Zhang-Server. In particular, when only the targets with NEFF≤3 are considered, our method outperforms HHpred2, MUSTER and Phyre2 by 41.2, 34.4 and 32.3%, respectively. When only the targets with NEFF≤4 are considered, our method outperforms HHpred2, MUSTER and Phyre2 by 8.5, 9.2 and 10.1%, respectively. When only the targets with NEFF≤3 and ≤4 are evaluated, our method is better than Zhang-Server by 20.8 and 3.2%, respectively. If we exclude the five easy targets6 (i.e. T0390, T0442, T0447, T0458 and T0471) from evaluation, then our method is better than Zhang-Server, HHpred2, MUSTER and Phyre2 by 10.5, 15.9, 14.5 and 18.0%, respectively, on the 11 hard targets with NEFF≤4. The performance of our method on low-homology targets is significant considering that our method is a pure single-template based threading method while Zhang-Server combines results from ~10 threading programs and also refines models extensively. Our new method is also better than our own RAPTOR++ program on low-homology targets. In CASP8, RAPTOR++ uses three in-house threading methods, a multiple-template method for easy targets and also a template-free method for hard targets.

# Chapter 4

# A multiple-template approach for protein threading

## 4.1 Introduction

Traditional protein threading method, including the methods described in Chapter 2 and 3, builds the 3D structure of a target protein sequence using a single template protein (Venter, Adams et al. 2001; Xu and Li 2003; Xu, Li et al. 2003; Zhou and Zhou 2005; Wu and Zhang 2008; Peng and Xu 2009). Along with many more solved protein structures deposited to PDB, it is more likely that a target protein without solved structure has more than one good template structures. Therefore, to make full use of the solved structures in PDB, we need to extend the classical single-template protein threading method so that a target protein sequence can be threaded onto multiple templates simultaneously and thus, its 3D model can be built from multiple template structures.

Template-based modeling may be improved using multiple templates in several aspects. First, it is very challenging to choose the best single template for a target protein when it has several similar templates in PDB. We can circumvent this challenging problem if we use multiple similar templates to build a 3D model for the target. Second, we can increase alignment coverage for the target protein using multiple templates (Cheng 2008; Larsson, Wallner et al. 2008). That is, we can align more regions in a target protein to the multiple templates than to a single template so that more regions in a target protein can be modeled. In addition, multiple templates may be complementary to one another in terms of their similarity to the target protein. That is, the target protein may be similar to one template in one region and to another template in another region. Therefore, we can improve modeling accuracy by copying structure information from the most similar template regions (Fernandez-Fuentes, Madrid-Aliste et al. 2007). Finally, we can also improve protein alignment accuracy through structural similarity among multiple templates. Alignment accuracy directly determines the quality of a template-based 3D model, so it is critical to generate an accurate sequence-template alignment. Existing multiple-template approaches (Fernandez-Fuentes, Madrid-Aliste et al. 2007; Joo, Lee et al. 2007; Cheng 2008; Larsson, Wallner et al. 2008; Rykunov, Steinberger et al. 2009) usually demonstrate that using multiple templates can improve alignment coverage of the target, but not alignment accuracy.

This chapter describes a novel probabilistic-consistency method that can align a single protein sequence simultaneously to multiple templates. We develop this method by extending our single-template threading method BoostThreader (see Chapter 2 and 3 and (Peng and Xu 2009; Peng and Xu 2010)). BoostThreader not only generates an accurate sequence-template alignment, but also efficiently calculates the (marginal) probability of one sequence residue being aligned to one template residue. Thus we can use a probabilistic alignment matrix to represent the alignment space of the sequence and each of its templates. Each entry in the matrix is the (marginal) alignment probability of two residues calculated from BoostThreader. Our multiple-template method generates the multiple sequence/template alignment by maximizing its probabilistic consistency with all the probabilistic alignment matrices. That is, two residues aligned in the multiple sequence/template alignment should have a high probability in their alignment matrix.

The probabilistic-consistency method has been used by ProbCons (Do, Mahabhashyam et al. 2005) for multiple sequence alignment. ProbCons cannot be directly used for multiple-template threading when proteins under consideration are distantly-related because 1) ProbCons does not use much evolutionary and structural information in generating a probabilistic alignment matrix; and 2) ProbCons ignores gap penalty since it is very expensive to estimate the probability of a gap. It is fine to ignore gap penalty when proteins to be aligned are close homologs. However, ignoring gap penalty deteriorates alignment accuracy when proteins under considerations are distantly-related. By contrast, our probabilistic-consistency method takes into consideration gap penalty so that we can handle distantly-related proteins. We achieve this by developing a novel approximation method that can accurately estimate the probability of a gap efficiently.

## 4.2 Methods

### *4.2.1 Overall algorithm*

As shown in **Algorithm 4.1**, the workflow of our multiple-template threading method is as follows. Given a target sequence, we first run our single-template method BoostThreader to determine the top templates of the target. Then we build a probabilistic alignment matrix for any two templates from their pairwise structure alignments generated by TMalign (Zhang and Skolnick 2005) and Matt (Menke, Berger et al. 2008). We also generate the probabilistic alignment matrix between the target and each template using BoostThreader. Afterwards, we run probabilistic-consistency transformation to iteratively update all the probabilistic alignment matrices. Finally, we generate a multiple sequence/template alignment by progressive alignment and refinement and run MODELLER (Fiser and Sali 2003) to build a 3D model from the alignment.

> **Input:** the query sequence
>
> 1. Run BoostThreader and get a set of top templates M.
>
> 2. Generate the probabilistic alignment matrix between the query and each template in M by dynamic programming.
>
> 3. Generate the probabilistic alignment matrix between any two templates in M by TMalign and Matt programs.
>
> 4. Repeat until convergence
>
>     a. Update the probabilistic alignment matrices according to our consistency transformation
>
> 5. Compute the initial multiple alignment by progressive alignment, maximizing the sum of pair scores
>
> 6. Repeat until convergence

**Algorithm 4.1.** The overall algorithm of our multiple-template threading method.

*4.2.2 A probabilistic-consistency transformation*

Given a set of proteins to be aligned, the key idea of the consistency method is to make their multiple alignment as consistent as possible with their pairwise alignments. Instead of fixing the alignment between two proteins, the probabilistic-consistency method uses a probabilistic alignment matrix to represent all the possible alignments between two protein residues; each alignment is associated with a probability. Then the probabilistic-consistency method will adjust the entries in the alignment matrices to achieve the maximum consistency among all the alignment matrices. Given two proteins x and y, let $P(x_i \circ y_j)$ denote the alignment probability of two residues $x_i$ and $y_j$. The probabilistic-consistency method adjusts the alignment probability between $x_i$ and $y_j$ through their alignments to an auxiliary protein z. If a residue $z_k$ in z aligns to both $x_i$ and $y_j$ with high probability, $x_i$ and $y_j$ are more likely to be aligned. We can calculate the alignment probability of $x_i$ and $y_j$ given z as follows.

$$P(x_i \circ y_j \mid z) = \sum_{z_k} P(x_i \circ y_j \circ z_k) + \sum_{z_{(k,k+1)}} P(x_i \circ y_j \circ z_{(k,k+1)}) \qquad (1)$$

In Equation (1), $P(x_i \circ y_j \circ z_k)$ is the alignment probability of three residues $x_i$, $y_j$ and $z_k$ and $P(x_i \circ y_j \circ z_{(k,k+1)})$ is the alignment probability of two residues $x_i$ and $y_j$ and a gapped position $z_{(k,k+1)}$ between the $k^{th}$ and $(k+1)^{th}$ residues. If we assume that the alignment between x and z is independent of that between y and z, we can decompose the first item in Equation (1) into the product of $P(x_i \circ z_k)$ and $P(y_j \circ z_k)$. Similarly, we can also decompose the second item in Equation (1) into a product of three items: $P(x_i \circ y_j)$, $P(x_i, z_{(k,k+1)})$ and $P(y_j, z_{(k,k+1)})$. It is challenging to estimate $P(x_i, z_{(k,k+1)})$ and $P(y_j, z_{(k,k+1)})$ since the probabilistic alignment matrices do not explicitly contain information relevant to gaps.

In order to estimate the second item in (1), we merge all the gapped positions in z into a single *GAP* state. Let $P(x_i \circ z_{GAP})(=1-\sum_k P(x_i \circ z_k))$ denote the probability of $x_i$ not being aligned to any residues in z. We can approximate the second item in Equation (1) as follows.

$$\sum_{z_{(k,k+1)}} P(x_i \circ y_j \circ z_{(k,k+1)}) = \sum_{z_{(k,k+1)}} P(x_i \circ y_j) P(x_i \circ z_{(k,k+1)}) P(y_j \circ z_{(k,k+1)})$$

$$= P(x_i \circ y_j) \sum_{z_{(k,k+1)}} P(x_i \circ z_{(k,k+1)}) P(y_j \circ z_{(k,k+1)})$$

$$\approx P(x_i \circ y_j) \sum_{z_{(k,k+1)}} P(x_i \circ z_{(k,k+1)}) \sum_{z_{(k,k+1)}} P(y_j \circ z_{(k,k+1)})$$

$$= P(x_i \circ y_j) P(x_i \circ z_{GAP}) P(y_j \circ z_{GAP})$$

That is, the second item in Equation (1) is approximated as the product of three terms: $P(x_i \circ y_j)$, $P(x_i \circ z_{GAP})$ and $P(y_j \circ z_{GAP})$. This approximation works well empirically. We can achieve very good alignment accuracy without incurring much more computational burden.

Treating all the templates of a target equally, we have the following probabilistic-consistency transformation formula,

$$P^{t+1}(x_i \circ y_j) \leftarrow \frac{1}{|M|} \sum_{z \in M} \left( \sum_k P^t(x_i \circ z_k) P^t(y_j \circ z_k) + P^t(x_i \circ y_j) P^t(x_i \circ z_{GAP}) P^t(y_j \circ z_{GAP}) \right)$$

where M is the set of available templates and $t$ is the number of iterations of probability adjustment. We can efficiently calculate $P^t(x_i \circ z_{GAP})$ and $P^t(y_j \circ z_{GAP})$ before each round of probabilistic-consistency transformation starts so that the second item in the above equation can be efficiently calculated.

We iteratively update the probabilistic alignment matrices until convergence or 20 iterations of probability adjustment are executed. Once the probabilistic-consistency transformation is finished, we will perform progressive alignment and iterative refinement to generate a multiple alignment.

### 4.2.3 Implementation details

***The probabilistic alignment matrix for a pair of target and template.*** Given a pair of target and template $(x, y)$, their probabilistic alignment matrix $P_{x,y}$ is computed using our single-template threading method BoostThreader as follows.

$$P(x_i \circ y_j) = P_{x,y}(i, j) = \sum_{a \in A} I(x_i \circ y_j \in a) P(a \mid x, y) ,$$

where $P_{x,y}(i, j)$ is the (marginal) alignment probability of residues $x_i$ and $y_j$; $A$ is the set of all possible alignments between x and y; $I(x_i \circ y_j \in a)$ is an indicator function, which equals to 1 if $x_i$ and $y_j$ are paired in the alignment $a$, otherwise 0; $P(a \mid x, y)$ is the probability of an alignment $a$ between x and y calculated from BoostThreader. The probabilistic alignment matrix can be efficiently computed using the forward-backward algorithm in O(MN) time where M and N are the lengths of sequence and template respectively.

*The probabilistic alignment matrix for two templates.* We construct a probabilistic alignment matrix between two templates using two structure alignment programs TMalign (Zhang and Skolnick 2005) and Matt (Menke, Berger et al. 2008). From a pairwise structure alignment, we build a binary matrix by setting the entry corresponding to two aligned residues with value 1. The probabilistic alignment matrix is the average of two binary matrices.

*Progressive alignment.* Given all the probabilistic alignment matrices, it is still NP-hard to calculate the optimal multiple sequence/template alignment maximizing the probabilistic consistency. The computational complexity is exponential with respect to the number of proteins to be aligned. We use a heuristic method, called progressive alignment, to generate a multiple sequence/template alignment (Feng and Doolittle 1987). The method first builds a guide tree, which represents the hierarchical relationship among proteins, and then builds the multiple protein alignment gradually. We use the same procedure as ProbCons to build the guide tree and the final multiple sequence/template alignment.

*Iterative refinement.* Progressive alignment cannot guarantee a globally optimal solution. Errors appearing in the early stage of the progressive alignment are likely to be propagated to the final result. We use an iterative refinement method to improve the quality of the alignment (Gotoh 1996). In the beginning of each refinement step, proteins under consideration are randomly partitioned into two subsets. Then a new alignment is constructed by aligning the alignments of these two subsets through maximizing the probabilistic consistency. In this work, we run 100 iterative refinement steps after progressive alignment.

***Selection of top templates.*** Given a target protein, BoostThreader first threads it to all the templates in the database PDB95. PDB95 is a set of representative proteins with solved structures and any two proteins in this set have less than 95% sequence identity. Afterwards, BoostThreader ranks all the templates using the neural network regression model described in Chapter 3, which predicts the quality (i.e., TM-score) of a target-template alignment (Peng and Xu 2009). A template is discarded if its alignment to the target has a predicted quality less than 90% of the best predicted quality. At most 20 templates are kept for further selection. The pairwise structure similarity between any two templates, measured by TM-score, is calculated using TMalign/Matt. A template is discarded if its structure similarity with the first-ranked template is low (e.g., TM-score<0.65) or less than 90% of the best predicted sequence-template alignment quality. By this way, we make sure that the target and its top templates are mutually similar and thus, a meaningful multiple alignment can be constructed among them.

***Computational complexity.*** The computational complexity of each round of probabilistic-consistency adjustment in our method is in the same order of magnitude as that of ProbCons. The total computational time of both ProbCons and our method is also linear with respect to the number of probability-consistency adjustment iterations. Since our method usually executes more rounds of probability adjustment to achieve the best alignment accuracy for a set of distantly-related proteins, it takes more but reasonable time for our method to terminate.

## 4.3 Experiments on CASP targets

To evaluate our multiple-template threading algorithm, we use a subset of 51 CASP8 targets and 48 CASP9 targets, all of which have at least two reliable templates. For each target, we determine its templates using our single-template threading program BoostThreader (Peng and Xu 2009; Peng and Xu 2010). Note that our results on the 48 CASP9 targets are directly taken from our RaptorX server submissions to CASP9. That is, these results were generated without knowing the native structures. We compare our method with other multiple sequence/structure alignment tools using reference-independent alignment accuracy, which is the most important measure in CASP to evaluate the performance of a method. The reference-independent alignment accuracy of an alignment is defined as the quality of the 3D model built from the alignment. To ensure a fair comparison, all these methods use the same set of templates for a given target. Given a target, we first align its sequence to its templates using the multiple-alignment methods and then use MODELLER 9v3 with default parameters (Sali 1995) to build 3D models from the multiple-alignments. We can evaluate the quality of a 3D model, measured by TM-score (Zhang and Skolnick 2004) and GDT-TS, by comparing the model with its native structure. Both TM-score and GDT-TS are two widely-used measures for model quality. TM-score ranges from 0 to 1 while GDT-TS from 0 to 100. The higher TM-score/GDT-TS, the better quality the model has. The native structures used for evaluation are downloaded from Zhang's CASP assessment website (http://zhanglab.ccmb.med.umich.edu/casp9/).

In Table 4.1, BoostThreader is our single-template threading method. Baseline is a naïve multiple-template method which simply assembles the BoostThreader pairwise sequence-template alignments into a multiple sequence/template alignment without correcting alignment errors through template structural similarity. The baseline method may result in larger alignment coverage than BoostThreader, but cannot correct alignment errors in BoostThreader. MAFFT (Katoh, Misawa et al. 2002), T-Coffee (Notredame, Higgins et al. 2000), MUSCLE (Edgar 2004) and ProbCons are multiple sequence alignment methods (i.e., no structure information is used). PROMALS3D (Pei, Kim et al. 2008) is a multiple sequence/structure alignment method. Both sequence profile and structural information is employed in PROMALS3D. M-Coffee is a meta-multiple alignment tool (Wallace,

O'Sullivan et al. 2006), which generates a multiple-alignment by combining pairwise structure alignments generated by TMalign (Zhang and Skolnick 2005), multiple structure alignment by Matt (Menke, Berger et al. 2008) and pairwise sequence-template alignments by BoostThreader. We also developed a new program ProbCons2, which uses the same procedure as our multiple-template threading method to generate probabilistic alignment matrices between two proteins, but uses the probabilistic-consistency procedure in ProbCons to generate the final multiple-alignment. By comparing our multiple-template threading method with ProbCons2 and M-Coffee, we can demonstrate the superiority of our probabilistic-consistency procedure.

**Table 4.1.** Cumulative TM-score and GDT-TS of the models generated by various multiple sequence/structure alignment methods. P-values in the table are calculated from a paired student t-test between our method and others. The smaller the P-value is, the more likely our method is better. See text for the description of the methods in this table.

|  | Model Quality Score | | P-value | |
| --- | --- | --- | --- | --- |
| Methods | TM-score | GDT-TS | TM-score | GDT-TS |
| **Our method** | **75.876** | **6598.2** | - | - |
| Baseline | 73.386 | 6353.4 | 2.49E-08 | 2.38E-07 |
| BoostThreader | 72.863 | 6265.7 | 3.57E-14 | 1.94E-17 |
| MAFFT | 66.368 | 5715.9 | 3.69E-10 | 4.40E-10 |
| T-coffee | 67.697 | 5852.1 | 1.07E-07 | 9.19E-08 |
| MUSCLE | 66.556 | 5715.3 | 1.91E-09 | 1.54E-09 |
| ProbCons | 67.193 | 5804.9 | 3.93E-08 | 3.89E-08 |
| PROMALS3D | 72.636 | 6309.2 | 1.47E-04 | 3.94E-04 |
| ProbCons2 | 73.553 | 6390.6 | 6.87E-04 | 1.80E-03 |
| M-coffee | 73.721 | 6414.9 | 2.91E-04 | 1.69E-03 |

***Multiple-template threading outperforms single-template threading.*** As shown in Table 4.1, the cumulative TM-score and GDT-TS of the models generated by our multiple-template threading method are 75.876 and 6598.2, respectively, which are better than our single-template method BoostThreader (72.863 and 6265.7, respectively). A paired student t-test indicates that our multiple-template method excels BoostThreader significantly with P-values 3.57E-14 and 1.94E-17, respectively. In fact, our multiple-template threading can generate better 3D models for 88 out of the 99 targets than BoostThreader. Note that our multiple-template threading method is built from BoostThreader. This indicates that using multiple templates can indeed improve modeling accuracy for most targets. In the above comparison, we use the first-ranked templates chosen by BoostThreader to build the single-template models. Even if we use the best template (among the templates used to build multiple-template models) to build the single-template model for each target, our multiple-template method still excels the single-template method with P-values 3.77E-07 and 1.24E-09, respectively. This implies that it is still worth to use multiple-template methods instead of single-template methods even if we have a perfect template selection procedure.

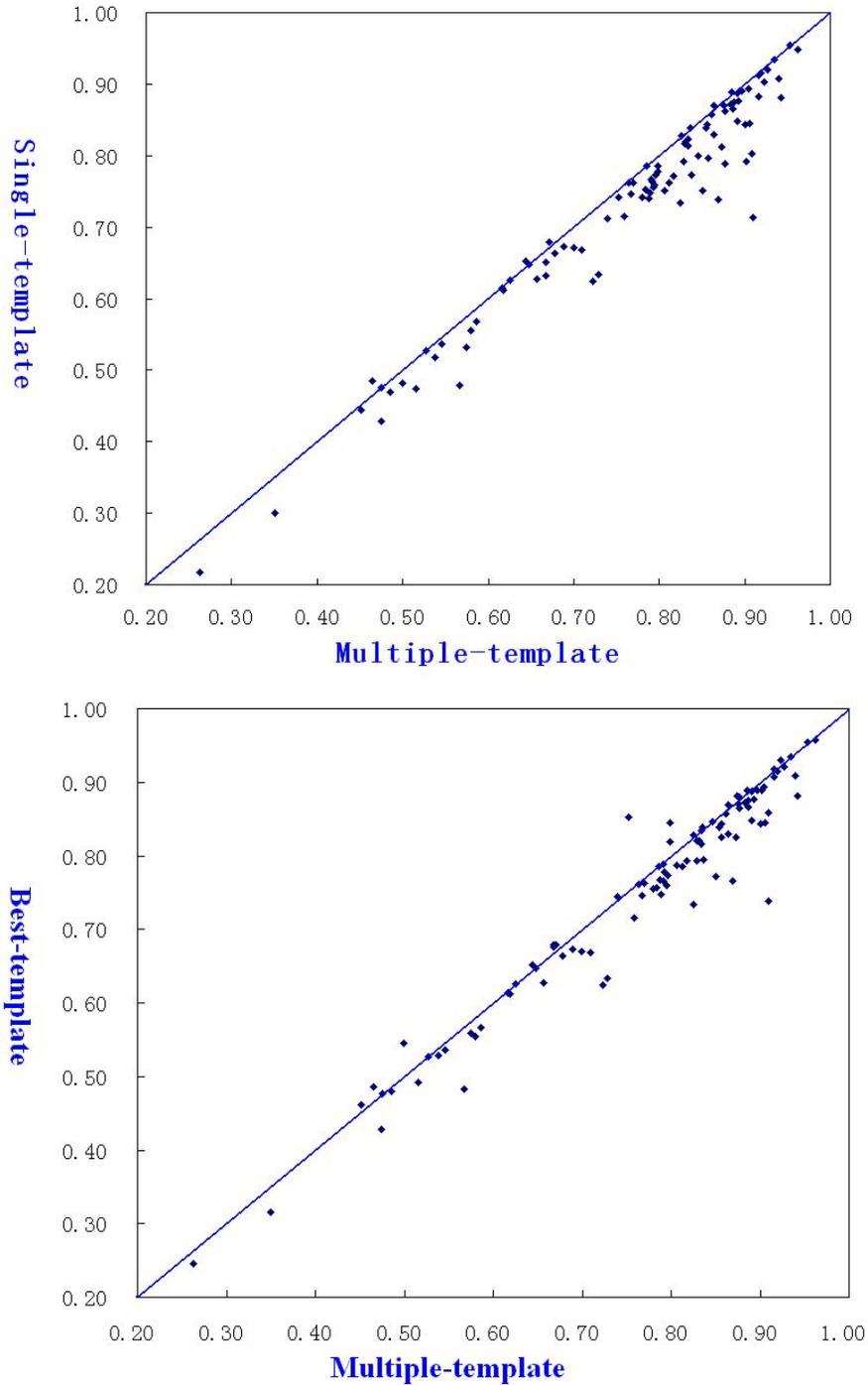

**Figure 4.1.** Our multi-template approach outperforms the single-template approach. The comparison betweeen BoostThreader and the multiple-template approach is shown. "Single-template" denotes the predictions from the first-ranked templates by BoostThreader. "Best-template" denotes the predictions from the best possible templates used in the multiplet-template threading.

*A consistent multiple sequence/template alignment is critical to model quality.* As shown in Table 4.1, our multiple-template method excels the baseline method significantly with P-values 2.49E-08 (TM-score) and 2.38E-07 (GDT-TS), respectively. In fact, the baseline method only performs marginally better than BoostThreader with P-values 0.2448 and 0.0645, respectively. These results indicate that using multiple templates (to increase alignment coverage) does not warrant an improvement in modeling accuracy unless we can generate a high-quality multiple sequence/template alignment, maybe because the benefit from increased alignment coverage is offset by the inconsistency and errors in the alignment.

*Our multiple-template method generates better alignments than other multiple sequence/structure alignment methods.* As shown in Table 4.1, our multiple-template method generates significantly better alignments than a bunch of popular multiple sequence alignment tools including MAFFT, T-Coffee, MUSCLE and ProbCons. This is expected since these tools do not use structure information and sequence profile in building alignments. Our method also outperforms several multiple sequence/structure alignment methods including PROMALS3D, M-Coffee and ProbCons2, all of which uses some "consistency" method to build multiple sequence/structure alignment. M-Coffee, ProbCons2 and our method all use BoostThreader to generate the (probabilistic) alignment matrix for a pair of sequence and template and TMalign/Matt to generate structure alignments among templates. This experimental result indicates that our "consistency" method is better than those used in M-Coffee and ProbCons for multiple sequence/structure alignment.

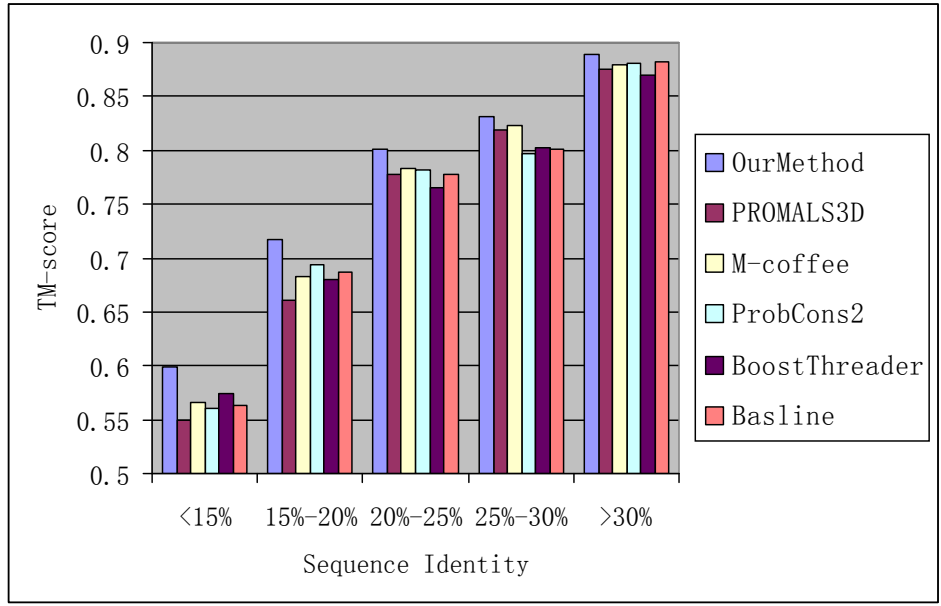

**Figure 4.2.** Average TM-score of the models for the targets in a group. The targets are divided into 6 groups according to their sequence identity to their best templates.

Our method performs especially well on distantly-related proteins. We divide the test targets into 5 groups according to their sequence identity to their best templates. In total, there are 15, 26, 29, 13, and 16 targets with sequence identity <15%, 15%-20%, 20%-25%, 25%-30% and >30% to their best templates, respectively. The average quality of the models for the targets in each group is calculated and shown in Figure 4.2. As shown in this figure, our method performs much better on the targets with low sequence identity to their best templates. We also observe the same trend when GDT-TS is used to evaluate model quality. When sequence identity is below 20%, PROMALS3D, M-Coffee and ProbCons2 even perform no better than our single-template method BoostThreader although both M-Coffee and ProbCons2 use BoostThreader to generate pairwise alignment (matrices). When sequence identity is below 20%, PROMALS3D is even worse than our single-template method BoostThreader although PROMALS3D uses both structure alignment and sequence profile to build multiple alignments.

***Our multiple-template threading method performs well in CASP9.*** Our multiple-template threading method is incorporated into our CASP9 server RaptorX for blind test. Overall, RaptorX is only slightly inferior to Zhang-Server (Zhang 2008; Zhang 2009) according to the assessment by Zhang group. On the set of 48 CASP9 targets with at least two reliable templates, our method obtained GDT-TS 3058.5. By contrast, the other five leading servers Zhang-Server (Zhang 2008; Zhang 2009), BAKER-ROBETTA (Das and Baker 2008; Raman, Vernon et al. 2009), HHpredA (Soding 2005), pro-sp3-TASSER (Zhou and Skolnick 2009) and Phyre2 (Kelley and Sternberg 2009) obtained GDT-TS 3075.5, 2796.5, 3029.3, 2883.9 and 2916.5, respectively. A paired student t-test also shows that our method excels Baker-Robetta, Phyre2 and pro-sp3-TASSER significantly ($p<0.001$) while the difference between our method and Zhang-Server ($p=0.827$) and HHpredA ($p=0.483$) is insignificant. Among these servers, Zhang-Server, BAKER-ROBETTA, and pro-sp3-TASSER refined their post-threading models extensively using computational-expensive folding simulation techniques with distance constraints extracted from multiple templates. Zhang-Server also uses a consensus method to choose the best templates from the outputs of ~10 threading programs. By contrast, our method can generate models with better or comparable accuracy without consensus or any refinement procedure and thus, our method is much more efficient. HHpredA is also a multiple-template method derived from HHpred (Soding 2005), but not published yet. Note that in the above comparison, the performance difference among servers may come from the choice of different templates for the same target.

***Comparison with ProbCons.*** Our probabilistic update is very similar to that is used in ProbCons for multiple sequence alignment. The major difference is that ProbCons ignores the second item in the right hand side of Equation (1) since ProbCons does not have an efficient method to estimate this item. It is fine to ignore this item when the following two conditions are satisfied: 1) proteins under consideration are close homologs since in this case the second item is much smaller than the first item; and 2) only a small number of iterations are executed to update the probabilistic alignment matrices. It is not very difficult to prove that if the second item in Equation (1) is ignored, then all the probabilistic alignment matrices will approach 0 when the number of probability-consistency iterations approaches to infinity. This is because at each round of probability adjustment, we will lose some alignment probability mass due to the loss of the second item in Equation (1). In the case we need to align a set of distantly-related proteins, we can neither ignore the second item, nor can we just update the probabilistic alignment matrices for a small number of iterations. Otherwise we cannot achieve the best alignment accuracy.

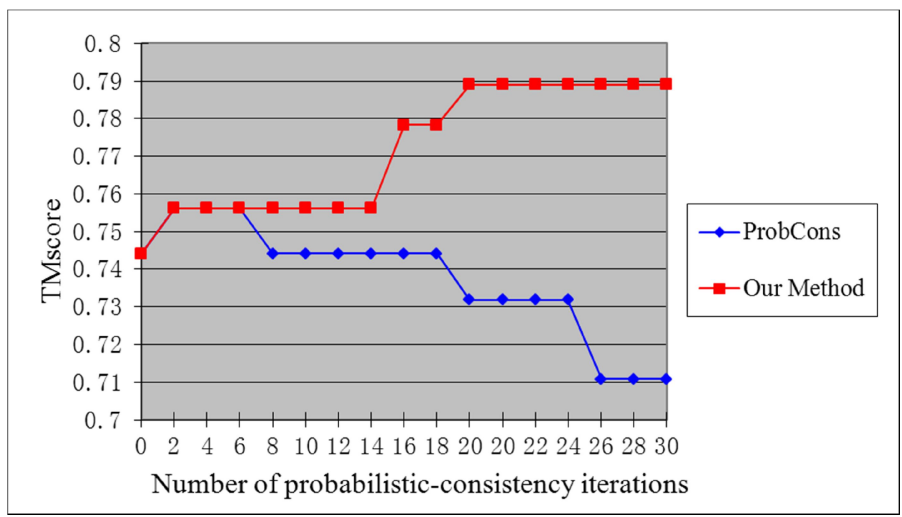

**Figure 4.3.** Comparison to ProbCons. The top panel shows the prediction accuracy of both methods as the number of updates increasing. The bottom panel shows the numerical stability of both methods as the number of updates increasing.

Experimental results confirm our analysis. As shown in Figure 4.3, when the number of probabilistic-consistency iterations is small (<6), both our method and ProbCons generate alignments with almost the same accuracy. However, when more than six rounds of probability adjustments are executed, ProbCons deteriorates the alignment dramatically while our method improves the alignment a lot. Note that in this experiment, ProbCons uses BoostThreader to generate the initial probabilistic alignment matrices, so the comparison shown in this figure is fair.

ProbCons fails to generate good alignments when more iterations of probabilistic-consistency transformation are executed because the probabilistic alignment matrices in ProbCons approach to zero too fast. There are two major reasons why PAM in ProbCons approaches to zero so fast. One is the underestimation of $\sum_{i,j}P(x_i \circ y_j \circ z_k)$ by assuming independence between x and y and the other is ignoring gap probability. Our method partially corrects the issue in ProbCons by not ignoring gap probability. To validate our analysis, we have randomly picked up some test examples and for each one we have calculated the sum of all the entries in all the probabilistic alignment matrices after each round of probabilistic-consistency transformation. Experimental results indicate that the sum in ProbCons goes to zero extremely fast. By contrast, the sum in our method decreases much more slowly, although it still decreases mainly due to the underestimation of $\sum_{i,j}P(x_i \circ y_j \circ z_k)$. This may indicate that ignoring gap probability causes a more serious issue than independence assumption. By the way, if we want to further improve alignment accuracy, we need a better estimation of $P(x_i \quad y_j \quad z_k)$ to further reduce or even avoid the decay of the probabilistic alignment matrices (i.e., we cannot assume x and y are totally independent), which is currently under investigation by our group.

***Specific examples.*** To showcase the detailed improvement of our approach, we selected several representative examples from our CASP9 submissions (Figure 4.4 – 4.7). These case clearly showed that the improvement comes from: 1) multiple templates provide more coverage than any single template; 2) many errors in the pairwise alignment have been corrected in the final multiple alignment.

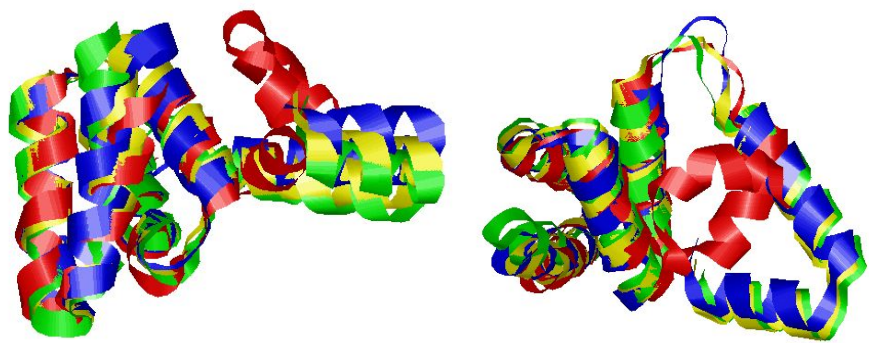

(a)

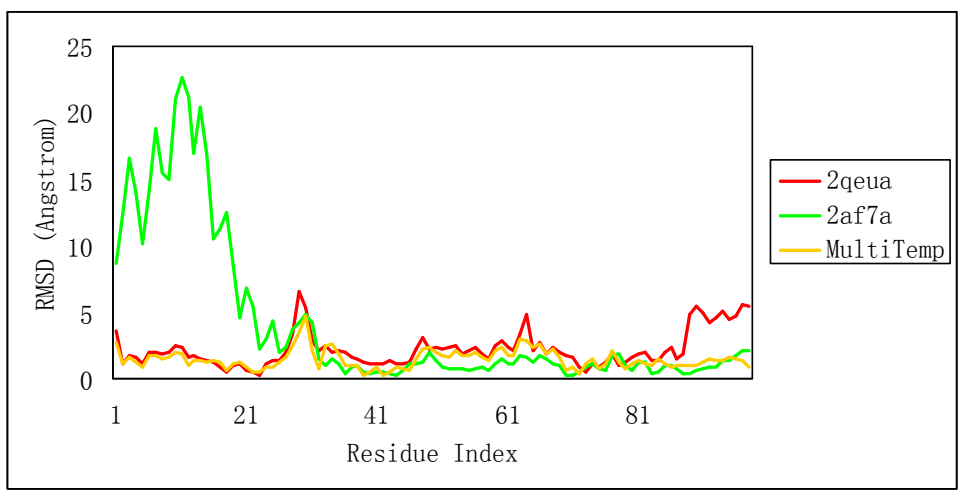

(b)

**Figure 4.4.** (a) Two different views of the structure superposition among the native (blue) of T0408, the model built from the best single template 2qeua (red, TM-score=0.73), the model built from 2af7a (green, TM-score=0.77) and the model built from both templates (yellow, TM-score=0.86). (b) The curves showing the per-position RMSD away from the native.

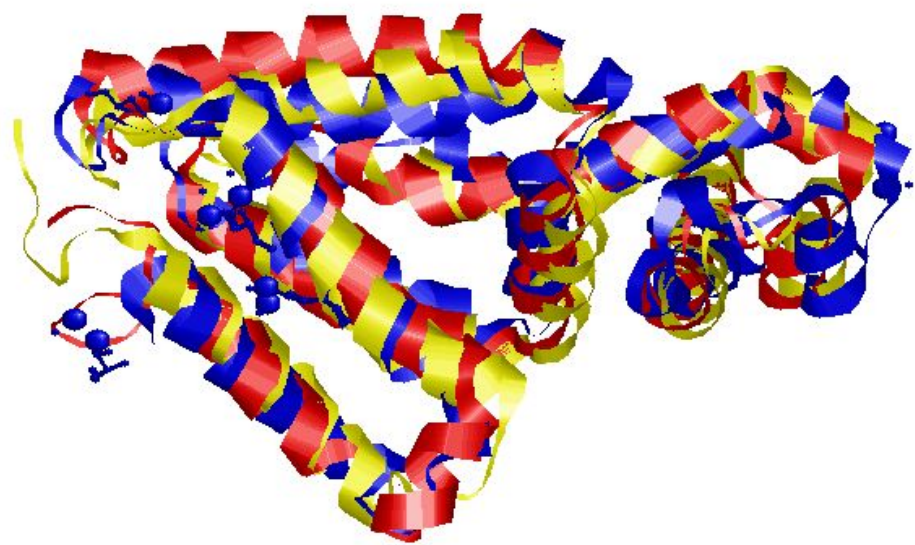

(a)

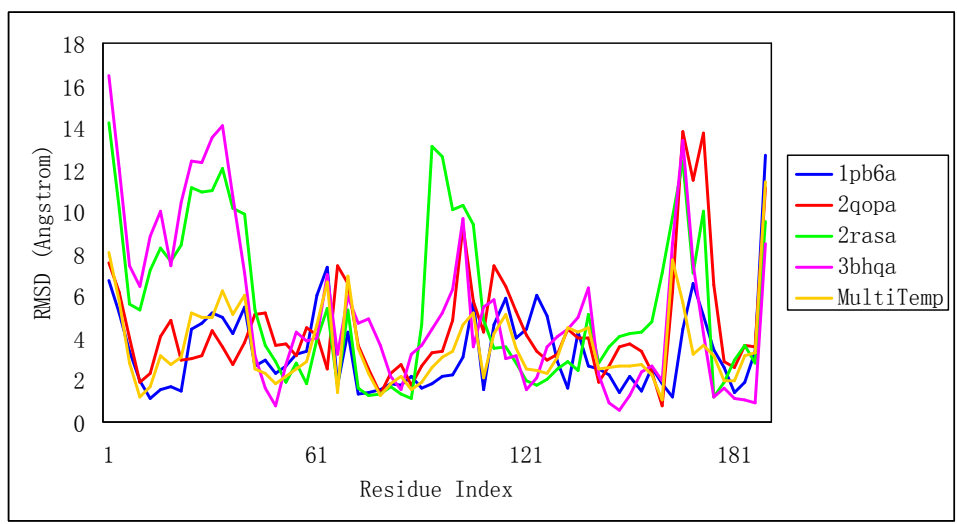

(b)

**Figure 4.5.** (a) The structure superposition among the native (blue) of T0454, the model built from the best single template 1pb6a (red, TM-score=0.71) and the model built from multiple templates including 1pb6a, 2qopa, 2rasa and 3bhqa (yellow, TM-score=0.76). (b) The curves showing the per-position RMSD away from the native.

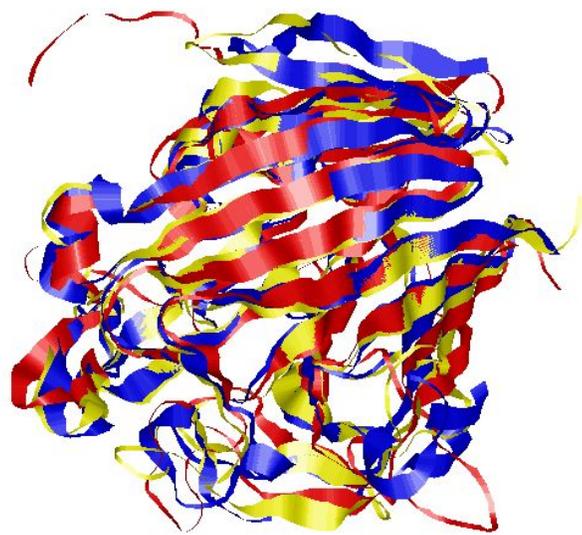

(a)

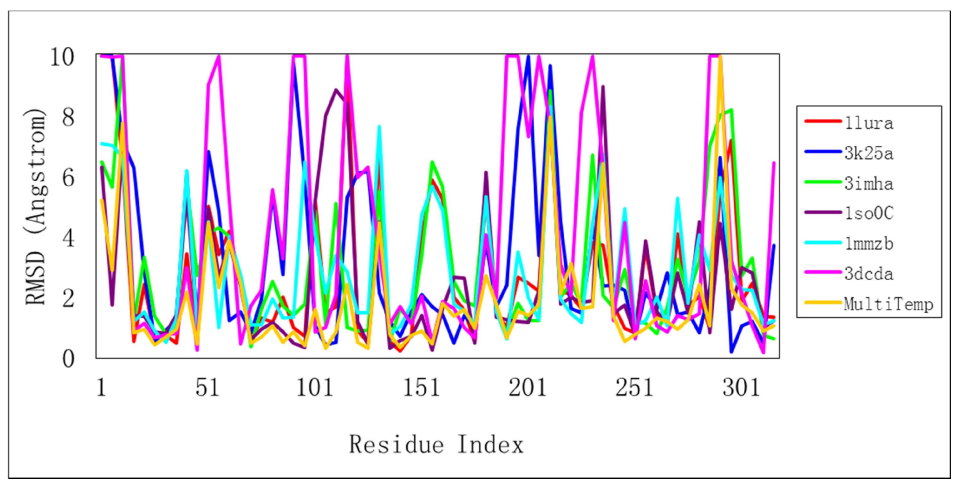

(b)

**Figure 4.6.** (a) The structure superposition among the native (blue) of T0524, the model built from the best single template 3k25a (red, TM-score=0.8) and the model built from multiple templates including 3k25a, 3dcda, 1lura, 3imha, 1so0c and 1mmzb (yellow, TM-score=0.91). (b) The curves showing the per-position RMSD away from the native. The RMSD at a position is set to 10Å if it is larger than 10Å so that the difference among models can be observed more easily.

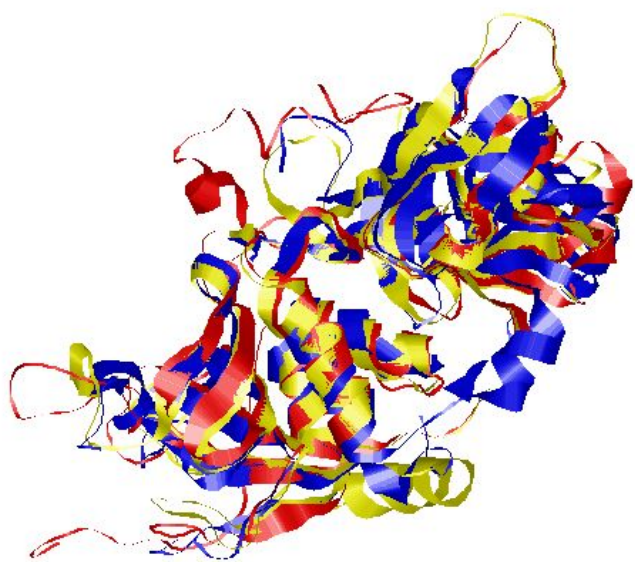

(a)

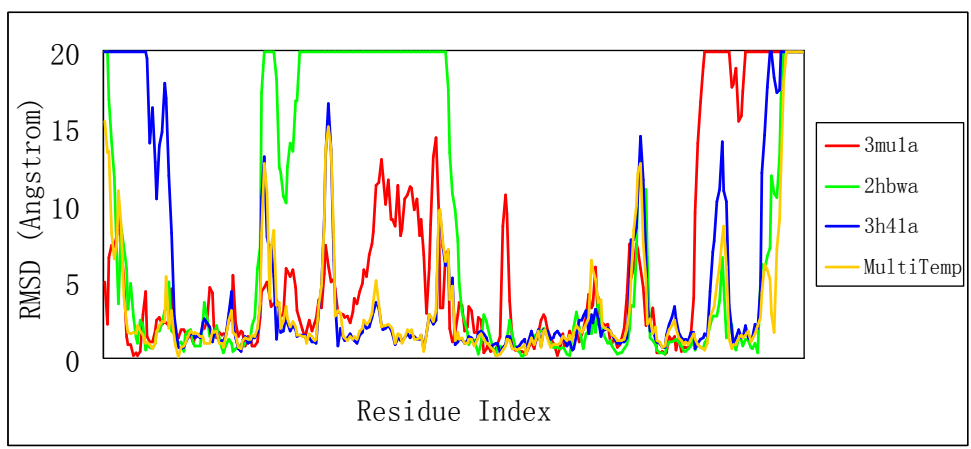

(b)

**Figure 4.7.** (a) The structure superposition among the native (blue) of T0565, the model built from the best single template 3h41a (red, TM-score=0.73) and the model built from multiple templates including 3h41a, 2hbwa and 3mu1a (yellow, TM-score=0.82). (b) The curves showing the per-position RMSD away from the native. The RMSD at a position is set to 20Å if it is larger than 20Å so that the difference among models can be observed more easily.

# Chapter 5

# Conclusion and Future Work

## 5.1 Summary of this thesis

Template-based modeling is probably the most successful approach for protein structure prediction to date. The main idea is to find the closest known structures to a query protein sequence, build alignments between the query and those, and use their structures as templates to build three-dimensional structures. The rationale behind this approach is that nearly all possible scaffolds or folds for protein structures in nature, about 1500 to 2000, are known, although the sequence space of all proteins is substantially larger. Most existing TBM methods rely on sequence alignment algorithms that additively consider evolutionary traces and structural features. Although these methods can make predictions with reasonable accuracy, many mistakes are made in identifying the best templates and building correct alignments. There is still a tremendously large gap between the predictive ability of template-based modeling methods and their theoretical limit. Furthermore, existing methods cannot fully exploit structural features, as well as the information from multiple templates. Instead of using a single template, a multiple template approach, better integrating structural features, could potentially improve the predictive power of template-based approaches. Supervised by Professor Jinbo Xu at TTI-C, I have been seeking to advance the state-of-the-art in TBM-based protein structure prediction along these lines.

First, we have applied a tree-based graphical model to probabilistically model pairwise protein alignment. Given two protein sequences, we constructed a Conditional Random Fields model for the pairwise alignment. In order to capture the complex dependencies between the evolutionary and structural similarities of protein sequences, we introduced a set of regression trees as the potential functions for this CRF model. This tree-based scoring function can be efficiently learned by maximizing the likelihood of a set of protein structure alignments through a functional gradient descent method. The fitted trees thus explain the relationship between evolutionary and structural signals. To better supervise the training of these regression trees, we have used an information-theoretical measure, NEFF, for each protein to quantify the strength of evolutionary signal implied by its homologous sequences. Guided by this feature, the trained model is able to automatically determine the importance of structural and evolutionary features according to the amount of homologous information. This is particularly important for proteins with sparse evolutionary signals, for which structural features become essential for prediction quality. Next, we proposed a probabilistic-consistency based approach to make use of multiple templates. Based on the CRF model for pairwise alignment, we designed an approximate but effective approach to construct a consensus alignment for multiple templates. The main idea is to minimize the inconsistency among all pairwise alignments. Notably, this approach not only constructs high-quality multiple alignments but also repairs many errors that appear in pairwise alignments, thus dramatically outperforming the single-template-based methods.

Together with my advisor Jinbo Xu, I have participated in recent CASP competitions to evaluate these methods, which are integrated into a RaptorX web server (raptorx.uchicago.edu). Our RaptorX method was ranked No. 2 in CASP9 and CASP10 in 2010 and 2012. Notably it achieved the best performance in the hard template-based modeling category. RaptorX was also voted by the CASP community as one of the most innovative methods. Since January 2012, the RaptorX webserver has predicted structures for more than 30,000 proteins submitted by more than 3,400 users from ~100 countries. This work was invited to both the CASP8 and CASP9 special issues at the journal Proteins; other technical publications from this work appeared in several top conferences including RECOMB, ISMB, NIPS and ICML, and several journals including Bioinformatics, Journal of Proteomics and Nature Protocols (Peng, Bo et al. 2009; Peng and Xu 2009; Xu, Peng et al. 2009; Peng and Xu 2010; Zhao, Peng et al. 2010; Peng and Xu 2011; Peng and Xu 2011; Wang, Zhao et al. 2011; Kallberg, Wang et al. 2012; Ma, Peng et al. 2012).

## 5.2 Future work

To go beyond the limitations of current alignment methods, a possible future direction is to design an alignment method that can better incorporate heterogeneous information from evolutionary and structural features. Conditional Neural Fields (CNFs) (Peng, Bo et al. 2009) are such a method that can potentially be applied to protein threading.

### *5.2.1 A Conditional Neural Fields model for protein threading*

CNFs are a recently developed probabilistic graphical model (Peng, Bo et al. 2009), which integrates the power of both Conditional Random Fields (Lafferty, McCallum et al. 2001) and neural networks. CNFs borrow from CRFs by parameterizing conditional probability in the loglinear form, and from neural networks by implicitly modeling complex, non-linear relationship between input features and output labels. CNFs have been applied to protein secondary structure prediction (Wang, Zhao et al. 2011), protein conformation sampling (Zhao, Peng et al. 2010) and handwriting recognition (Peng, Bo et al. 2009). The major advantage of CNFs over tree-based CRFs is that neural networks provide a more efficient nonlinear feature composition whose parameters can be estimated accurately through gradient-based training, while tree-based CRFs often require a large number of regression trees thus making the both training and inference very slow. Here we propose to model protein sequence-template alignment using a CNF model.

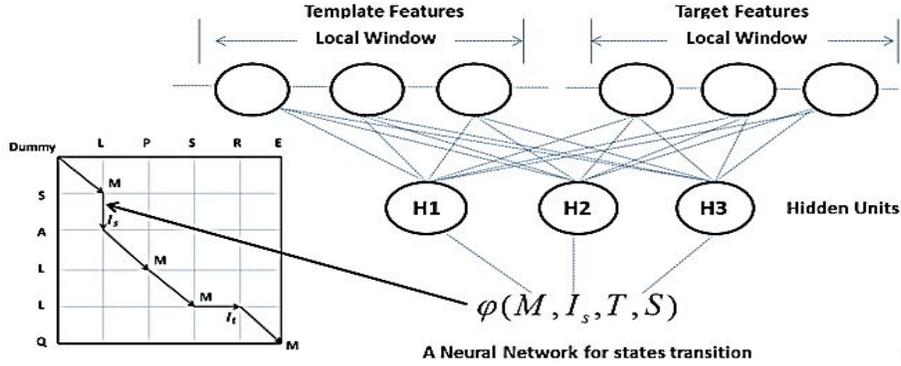

**Figure 5.1.** An example of the edge feature function, which is a neural network with one hidden layer. The function takes both template and target protein features as input and yields one likelihood score for state transition M to Is. Meanwhile, H1, H2 and H3 are hidden neurons conducting nonlinear transformation of the input features.

Similar to the CRF formulation proposed in Chapter 2, we calculate the probability of one alignment A as follows.

$$P(A|T,S,_n) = \exp(\sum_{i=1}^{L_A} E(a_{i-1}, a_i, T, S) / Z(T,S))$$

where $\theta$ is the model parameter vector to be trained, i indicates one alignment position and is the normalization factor (i.e., partition function) summing over all possible alignments for a given protein pair. The function E the above definition estimates the log-likelihood of state transition from $a_{i-1}$ to $a_i$ based upon protein features. It is a nonlinear scoring function defined as the sum of edge and node feature functions.

$$E(a_{i-1}, a_i, T, S) = \{ (a_{i-1}, a_i, T, S) + w(a_i, T, S)$$

Both the edge and label feature functions can be as simple as a linear function in CRFs or as complex as a neural network in CNFs. Here we use neural networks with only one hidden layer to construct these two types of functions. Because the label feature function can be incorporated into edge feature functions and slightly simpler, we only explain the edge feature function in detail. Since in total there are 9 possible state transitions in an alignment, we need 9 edge feature functions, each corresponding to one kind of state transition. Figure 5.1 shows an example of the edge feature function for the state transition from M to It. Given one state transition u to v at position i where u and v are two alignment states, the edge feature function is defined as follows.

$$\{(a_{i-1} = u, a_i = v, T, S) = \sum_j \}_{u,v}^j H_{u,v}^j (w_{u,v}^j f_{u,v}(T,S,i))$$

where function f is the feature vector, which consists of input features from the target and template proteins for the alignment at position i. The feature vector f is state-dependent, so we may use different features for different state transitions. j is the index of the hidden neurons in the hidden layer, $\lambda_{u,v}^j$ is the model parameter between one hidden neuron and the output layer, H is the logistic sigmoid gate function for the hidden neuron conducting nonlinear transformation of input, and $w_{u,v}^j$ is the model parameter vector connecting the input layer to one hidden neuron. All the model parameters are state-dependent, but position-independent. In total there are 9 different neural networks for the 9 state transitions. These neural networks have separate model parameters.

## 5.2.2 Alignment quality-sensitive training

CRFs/CNFs are usually trained by maximum likelihood (ML) or maximum a posteriori (MAP) (Volkovs and Zemel 2009). The ML method trains the CRFs/CNFs model parameters by maximizing the occurring probability of a set of reference alignments, which are built by a structure alignment tool. The ML method treats all the aligned positions equally, ignoring the fact that some are more conserved than others. It is important to align the conserved residues correctly since they may be related to protein function. As such, it makes more sense to treat conserved and non-conserved residues separately. Although there are a few measures for the degree of conservation to be studied, here we simply use the local TM-score (Zhang and Skolnick 2004) between two aligned residues. Given a reference alignment (and the superimposition of two proteins in the alignment), the local TM-score at one alignment position *i* is defined as follows.

$$w_i = \frac{1}{1 + (d_i / d_0)^2}$$

where $d_i$ is the distance deviation between the two aligned residues at position *i* and $d_0$ is a normalization constant depending on only protein length. TM-score ranges from 0 to 1 and the higher the more conserved the aligned position is. When the alignment state at position *i* is gap, the local TM-score is equal to 0 and $w_0$ is equal to 0 at a gap position. To differentiate the degree of conservation in the alignment, we train the CNF model by maximizing the expected TM-score. The central problem is to calculate the gradient of the following objective function:

$$Q = \frac{1}{N(A)} \sum_i (w_i MAG_i)$$

where *N(A)* is the length of the smaller protein in the alignment *A*, $w_i$ and $MAG_i$ are the local TM-score and marginal alignment probability at alignment position *i*, respectively.

***Notations.*** Given an alignment $A = \{a_1, a_2, ......, a_{L_A}\}$, let $A[1,i] = \{a_1, a_2, ..., a_i\}$ denote a left partial alignment starting from the N-terminal to position *i* and $A[i, L_A] = \{a_i, a_{i+1}, ..., a_{L_A}\}$ denote a right partial alignment starting from the C-terminal to position *i*. Let *x* and *y* denote the number of target and template residues contained in the left partial alignment *A[1,i]*, respectively. Both *x* and *y* can also be treated as the residue indices in the target and template proteins, respectively. Therefore, each alignment position index *i* is associated with a pair of residue indices *x* and *y*. Let *m* and *n* denote the number of residues in the target and template proteins, respectively. In total there are *mn* possible pairs of residue indices. Note that when alignment position *i* corresponds to a pair of residue indices *x* and *y*, the alignment position *i-1* may correspond to one of the three possible residue index pairs *(x-1,y-1)*, *(x,y-1)* or *(x-1,y)*, depending on the alignment state at position *i*.

***Gradient calculation.*** Let $F_i^v$ denote the accumulative probability of all possible left partial alignments ending at alignment position *i* with state *v*. Similarly, let $B_i^u$ denote the accumulative probability of all possible right partial alignments ending at alignment position *i* with state *u*. $F_i^v$ and $B_i^u$ are the forward and backward functions, respectively, which have been discussed in Chapter 2 for CRF training. Sometimes we also write $F_i^v$ as $F_{x,y}^v$ or $B_i^u$ as $B_{x,y}^u$ when it is necessary to explicitly spell out the residue indices. Both $F_i^v$ and $B_i^u$ can be calculated recursively as follows.

$$F_i^v = \sum_u F_{i-1}^u \exp(E(a_{i-1} = u, a_i = v, S, T))$$
$$B_i^u = \sum_v F_{i+1}^v \exp(E(a_{i-1} = u, a_i = v, S, T))$$

The marginal alignment probability $MAG_i$ can be calculated as follows.

$$MAG_i = \frac{F_i^M B_i^M}{Z}$$

Meanwhile, the normalization factor $Z$ (i.e., partition function) is equal to $\sum_u F_i^u B_i^u$ for any $i$. In particular, we have

$$Z = \sum_u F_{m,n}^u = \sum_u B_{1,1}^u$$

Since only $MAG_i$ depends on the model parameter $\theta$, so we only need to calculate $\frac{\partial MAG_i}{\partial_n}$ in order to calculate the gradient.

$$\frac{\partial MAG_i}{\partial_n} = \frac{\partial}{\partial_n}(\frac{F_i^M B_i^M}{Z}) = \frac{\partial F_i^M}{\partial_n}\frac{B_i^M}{Z} + \frac{\partial B_i^M}{\partial_n}\frac{F_i^M}{Z} - \frac{F_i^M B_i^M}{Z}\frac{\partial Z}{\partial_n}$$

Since $Z = \sum_u F_{m,n}^u$, we have $\frac{\partial Z}{\partial_n} = \sum_u \frac{\partial F_{m,n}^M}{\partial_n}$. That is, $\frac{\partial MAG_i}{\partial_n}$ can depend on only $\frac{\partial F_i^u}{\partial_n}$ and $\frac{\partial B_i^u}{\partial_n}$. For the purpose of simplicity, let $E_i^{u \to v}$ denotes $E(a_{i-1}=u, a_i=v, S, T)$. We have

$$\frac{\partial F_i^u}{\partial_n} = \sum_u (\frac{\partial F_{i-1}^u}{\partial_n}\exp(E_i^{u \to v}) + \exp(E_i^{u \to v})\frac{\partial E_i^{u \to v}}{\partial_n}F_{i-1}^u)$$

This equation indicates that $\frac{\partial F_i^u}{\partial_n}$ can be calculated recursively. Similarly, $\frac{\partial B_i^u}{\partial_n}$ can also be calculated recursively. Since $E_i^{u \to v}$ is a neural network, $\frac{\partial E_i^{u \to v}}{\partial_n}$ can be calculated using the gradient chain rule with time complexity depending on the architecture of the neural network. The size of the neural network is determined by the number of features, the window size and the number of hidden neurons, but independent of protein length. There are in total $mn$ possible residue index pairs for the alignment position $i$ in $F_i^v$ and $B_i^u$, so the time complexity of the gradient calculation is $O(mn)$, i.e., the product of the target and template protein lengths, If we assume the size of the neural network is a constant.

*5.2.3 Summary*

The presented protein threading method, which achieves much more accurate sequence–template alignment by employing a probabilistic graphical model called a Conditional Neural Field (CNF), aligns one protein sequence to its remote template using a non-linear scoring function. This scoring function accounts for correlation among a variety of protein sequence and structure features, makes use of information in the neighborhood of two residues to be aligned, and is thus much more sensitive than the widely used linear or profile-based scoring function. This CNF threading model can be trained with a a novel quality-sensitive method, instead of the standard maximum-likelihood method, to maximize directly the expected quality of the training set. This method can also be adapted to protein sequence alignment. Preliminary experimental studies of this approach can be found in a recently published article (Ma, Peng et al. 2012).